\documentclass{aa}

\usepackage{txfonts}
\usepackage{sistyle}
\usepackage{multirow}
\usepackage{graphicx,epstopdf}
\usepackage{hyperref}
\usepackage{natbib}
\bibpunct{(}{)}{;}{a}{}{,}
\DeclareMathOperator{\e}{e}
\graphicspath{{Images/}}

\newcommand{\ebv}{\ifmmode E_{\rm B-V} \else $E_{\rm B-V}$\fi}
\newcommand{\oh}{\ifmmode 12+\log({\rm O/H}) \else$12+\log({\rm O/H})$\fi}
\newcommand{\nh}{\ifmmode N_{\rm H~I} \else $N_{\ion{H}{i}}$\fi}
\newcommand{\fesc}{\ifmmode f_{\rm esc} \else $f_{\rm esc}$\fi}
\newcommand{\lya}{\ifmmode {\rm Ly}\alpha \else Ly$\alpha$\fi}
\newcommand{\sone}{SGAS J1226} 
\newcommand{\stwo}{SGAS J1527} 
\newcommand{\megasaura}{M\textsc{eg}a\textsc{S}a\textsc{ura}}

\begin{document}

   \title{Neutral gas properties of Lyman continuum emitting galaxies: Column densities and covering fractions from UV absorption lines}

   \author{S. Gazagnes\inst{1,2,3,4} \and J. Chisholm\inst{1} \and D. Schaerer\inst{1,5} \and A. Verhamme\inst{1} \and J. R. Rigby\inst{6} \and M. Bayliss\inst{7}}

 \offprints{s.r.n.gazagnes@rug.nl}
  \institute{
Observatoire de Gen\`eve, Universit\'e de Gen\`eve, 51 Ch. des Maillettes, 1290 Versoix, Switzerland
         \and
         Johan Bernouilli Institute, University of Groningen, P.O Box 407, 9700 Groningen, AK, The Netherlands 
   \and
   Kapteyn Astronomical Institute, University of Groningen, P.O Box 800, 9700 AV Groningen, The Netherlands
   \and
   KVI-Center for Advanced Radiation Technology (KVI-CART), Zernikelaan 25, Groningen 9747 AA, The Netherlands
         \and
 CNRS, IRAP, 14 Avenue E. Belin, 31400 Toulouse, France
 \and 
 Observational Cosmology Lab, NASA Goddard Space Flight Center, 8800 Greenbelt Rd., Greenbelt, MD 20771, USA
 \and 
 MIT Kavli Institute for Astrophysics and Space Research, 77 Massachusetts Ave., Cambridge, MA 02139, USA}

   \date{Received <date> /
Accepted <date>}

  \titlerunning{Neutral gas properties of Lyman continuum emitters}
  \abstract
    {The processes allowing the escape of ionizing photons from galaxies into the intergalactic medium are poorly known.}
  {To understand how Lyman continuum (LyC) photons escape galaxies, we constrain the \ion{H}{i} covering fractions and column densities using ultraviolet (UV) \ion{H}{i} and metal absorption lines of 18 star-forming galaxies that have Lyman series observations. Nine of these galaxies are confirmed LyC emitters}
  {We fit the stellar continuum, dust attenuation, metal, and \ion{H}{i} properties to consistently determine the UV attenuation, as well as the column densities and covering factors of neutral hydrogen and metals. We used synthetic interstellar absorption lines to explore the systematics of our measurements. 
  Then we applied our method to the observed UV spectra of low-redshift and $z\sim2$ galaxies.} 
 {The observed  \ion{H}{i} lines are found to be saturated in all galaxies. An indirect approach using \ion{O}{i} column densities and the observed O/H abundances yields \ion{H}{i} column densities  of $\log(\nh) \sim 18.6-20$ cm$^{-2}$. These columns are too high to allow the escape of ionizing photons. We find that the known LyC leakers have \ion{H}{i} covering fractions less than unity. Ionizing photons escape through optically thin channels in a clumpy interstellar medium. Our simulations confirm that the \ion{H}{i} covering fractions are accurately recovered. The \ion{Si}{ii} and \ion{H}{i} covering fractions scale linearly, in agreement with observations from stacked Lyman break galaxy spectra at $z \sim 3$. Thus, with an empirical correction, the \ion{Si}{ii} absorption lines can also be used to determine the  \ion{H}{i} coverage. Finally, we show that a consistent fitting of dust attenuation, continuum, and absorption lines is required to properly infer the covering fraction of neutral gas and subsequently to infer the escape fraction of ionizing radiation.}
{These measurements can estimate the LyC escape fraction, as we demonstrate in a companion paper.}

   \keywords{galaxies: ISM -- ISM: abundances -- ISM: lines and bands -- Ultraviolet: ISM -- dust, extinction -- dark ages, reionization, first stars }
   \maketitle

\section{Introduction}
Star-forming galaxies are ideal laboratories to understand how the early universe became reionized. Galaxies likely reionized the universe because quasars are too rare at high redshifts \citep{fontanot2012, fontanot2014}. Compact galaxies with intense star formation rates produce large amounts of ionizing photons which, under certain circumstances, escape the interstellar medium (ISM) and ionize the intergalactic medium (IGM). To reionize the universe, studies suggest that $10-20$\% of the ionizing photons produced by star-forming galaxies must escape galaxies \citep{ouchi, robertson, dressler}.  However, it has been challenging to detect Lyman continuum (LyC) radiation from individual galaxies.

Three unambiguous observations of ionizing photons have been reported at z\textasciitilde 3 \citep{vanzella2015,debarros2016,shapley2016,bian2017}. The observed escape fractions (near 50 \%) are greater than those required to reionize the universe. Additionally, there are nine low-redshift (z < 0.3)  galaxies with 1\% $ \leq f^{LyC}_{esc} \leq $ 13\% \citep{bergvall2006, leitet2013, borthakur2014, leitherer2016, izotov2016a, izotov2016b, puschnig2017}, and one recent detection at z = 0.37 with $f^{LyC}_{esc} \approx 46\%$  \citep{izotov2018}. The low number of detections emphasizes the difficulty of detecting Lyman continuum emitters (LCEs). 

\citet{zackrisson2013} proposed two theoretical models to explain how ionizing photons escape galaxies. In the first scenario, low \ion{H}{i} column densities allow Lyman continuum photons to pass through without being completely absorbed; this is called the density-bounded scenario \citep{jaskot2013, nakajima2014}. This scenario manifests itself as low \ion{H}{i} column densities ($<10^{18}$~cm$^{-2}$). In the second scenario, ionizing photons leak into the IGM through holes in the neutral gas \citep{heckman2011}. This scenario is called the picket-fence model and relies on a patchy neutral gas. A patchy neutral ISM manifests as \ion{H}{i} absorption lines with a covering fraction less than one. It is unclear which of these scenarios describes how ionizing photons leak from galaxies. Constraining neutral gas properties, especially the \ion{H}{i} covering fraction and column density, is an effective way to disentangle how ionizing photons escape galaxies.

\ion{H}{i} absorption lines in the rest-frame far ultraviolet (the Lyman series: 912-1026\AA) directly probe the \ion{H}{i} covering fraction and column density. However, the Lyman series is challenging to observe for several reasons. First, this requires deep rest-frame far-ultraviolet observations blueward of Ly$\alpha$, which is notoriously difficult to observe at low redshifts. Second, the Lyman series is unavailable at redshift z $>$ 3 because the Ly$\alpha$ forest completely absorbs this region. Third, several observational obstacles need to be accounted for to measure the \ion{H}{i} properties. In particular, foreground contamination \citep{vanzella2010}, intervening absorbers, stellar continuum from the galaxy itself, ISM absorption lines, and, at low redshifts, Milky Way and geocoronal emission need to be identified. 

Interstellar medium metal absorption lines (i.e., \ion{Si}{ii}~1260\AA, \ion{C}{ii}~1334\AA) are easier to observe than the Lyman series \citep{heckman2011, alexandroff2015}. However, using metal absorption lines to trace the \ion{H}{i} assumes that metals directly probe the neutral gas. Recent studies indicate that ISM metal lines may have a factor of 2 times smaller covering fractions than \ion{H}{i} absorption lines \citep{reddy2016stack}. As a result, metal absorption lines may not trace the \ion{H}{i}. 

In this article, we directly observe the Lyman series of individual high and low-redshift star-forming galaxies to determine their neutral gas properties, and to compare \ion{H}{i} measurements to ISM metal properties. For the first time, we measure the \ion{H}{i} properties of spectroscopically confirmed LyC emitters to determine which physical process enables ionizing photons to escape galaxies. A companion paper uses these observed \ion{H}{i} properties to predict the escape fractions of ionizing photons \citep[][hereafter Paper~II]{chisholm2018}.

This paper is organized as follows: Sect.~\ref{sect:dataobs} describes the observational data. Sect.~\ref{sect:method} defines the various models and equations used to fit the stellar continua and UV absorption lines. In Sect.~\ref{sect:sims} we use synthetic spectra to illustrate how accurately we recover \ion{H}{i} column densities and covering fractions from observations. Sect.~\ref{sect:results} discusses the measured \ion{H}{i} covering fractions of the LyC emitters, the relation between the \ion{H}{i} and \ion{Si}{ii} covering fractions, the effects of the assumed dust geometry, and comparisons to previous studies. We summarize our results in Sect.~\ref{sect:conc}.

\section{Observed data}
\label{sect:dataobs}
 
We studied the neutral gas properties of a sample of 18 star-forming galaxies listed in Table~\ref{table:sample}. Our selection was driven by the need to observe the Lyman series, i.e.,\ available rest-frame UV spectroscopy between Lyman-$\beta$ and the Lyman limit. We selected the low-redshift galaxies observed with the Cosmic Origins Spectrograph (COS) on the {\it Hubble Space Telescope} (HST) \citep{green2012} with this wavelength coverage.  Given the sensitivity and wavelength range of the G130M grating on COS, the Lyman series is observable with a spectral resolution $R \ga 1500$ for galaxies at $z>0.18$. Therefore, we selected the 15 galaxies at low redshifts ($z < 0.36$) with Lyman series observations from the COS archive. Each galaxy is at a redshift such that at least the Ly$\beta$ line is observable with the COS G140L or G130M gratings. 

Many of our low-redshift galaxies were originally targeted to observe possible LyC emission, but only nine of them are confirmed LyC emitting galaxies. Four galaxies are from the Green Pea sample of \citet{henry2015} and two are Lyman break analogs from \citet{heckman2011}. The other nine galaxies are known LyC emitters \citep[J1503+3644, J0925+1409, J1152+3400, J1333+6246, J1442-0209, J0921+4509,  Tol1247-232, Tol0440-381, and Mrk~54 from][]{izotov2016a, izotov2016b, borthakur2014, leitherer2016}. The nine leakers with COS/HST data were reduced using {\small CALCOS} v2.21 and a custom method for faint COS spectra \citep{worseck2016}. The other COS/HST data were reduced with {\small CALCOS} v2.20.1 and the methods from \citet{wakker2015}. The five \citet{izotov2016b, izotov2016a} spectra were smoothed with a 5 pixel boxcar.

Finally, we also included three gravitationally lensed galaxies at $z\approx3$ \citep{stark2008, koester}. These galaxies are part of the Magellan Evolution of Galaxies Spectroscopic and Ultraviolet Reference Atlas \citep[\megasaura;][]{rigby} and were selected because they are the only galaxies in the \megasaura\ sample with a signal-to-noise ratio (S/N) greater than 2 near the Lyman series. These are moderate resolution ($R \sim 3000$) spectra observed with the MagE spectrograph on the Magellan Telescopes \citep{marshall2008}. 
Instead of the full galaxy names, we used the following short names for the two sources: \sone~=~SGAS J122651.3+215220  and \stwo~=~SGAS J152745.1+065219.

The complete sample is summarized in Table~\ref{table:sample}, where we list a few host properties and (nominal) spectral resolutions of the observations. An upper limit on the [N II]/H$\alpha$ ratio constrains the metallicity for \stwo\ \citep[12+log(O/H) < 8.5;][]{wuyts}; for \sone\ these lines are not accessible from the ground. Meanwhile, \citet{stark2008} measured the metallicity of the Cosmic Eye using the R$_{23}$ index.
    \begin{table}
     \caption{Sample of galaxies with Lyman series observations} 
     \label{table:sample}
    \centering   
    \begin{tabular}{llllll}
    \hline \hline
    Galaxy name&  $z$ & \oh\ & $f_{esc}^{LyC}$ & $R$ \\
       (1) & (2) & (3) & (4) & (5) \\\hline
    J0921+4509  & 0.23499 & 8.67\tablefootmark{a} & 0.010\tablefootmark{g} & 15000  \\ 
    J1503+3644  & 0.3537  & 7.95\tablefootmark{b} & 0.058\tablefootmark{b} & 1500  \\ 
    J0925+1409  & 0.3013  & 7.91\tablefootmark{c} & 0.072\tablefootmark{c} & 1500  \\ 
    J1152+3400  & 0.3419  & 8.00\tablefootmark{b} & 0.132\tablefootmark{b} & 1500 \\ 
    J1333+6246  & 0.3181  & 7.76\tablefootmark{b} & 0.056\tablefootmark{b} & 1500  \\ 
    J1442-0209  & 0.2937  & 7.93\tablefootmark{b} & 0.074\tablefootmark{b} & 1500  \\ 
    Tol1247-232 & 0.0482  & 8.10\tablefootmark{d} & 0.004\tablefootmark{h} & 1500 \\
    Tol0440-381 & 0.0410  & 8.20\tablefootmark{d} & 0.019\tablefootmark{h} & 1500 \\
    Mrk54       & 0.0451  & 8.60\tablefootmark{d} & <0.002\tablefootmark{h} & 1500 \\
    J0926+4427  & 0.18069 & 8.01\tablefootmark{e} & -                      & 15000 \\ 
    J1429+0643  & 0.1736  & 8.20\tablefootmark{e} &  -                     & 15000  \\ 
    GP0303-0759 & 0.16488 & 7.86\tablefootmark{e} & -                      & 15000  \\ 
    GP1244+0216 & 0.23942 & 8.17\tablefootmark{e} & -                      & 15000  \\ 
    GP1054+5238 & 0.25264 & 8.10\tablefootmark{e} & -                      & 15000  \\ 
    GP0911+1831 & 0.26223 & 8.00\tablefootmark{e} &  -                     & 15000  \\
    \sone\      &  2.92525 & -                    &  -                     & 4000  \\ 
    \stwo\      &  2.76228 & $<8.5$ \tablefootmark{i}  &  -                & 2700 \\     
    Cosmic Eye  &  3.07483 & 8.60\tablefootmark{f}& -                  & 2500  \\ \hline
    \end{tabular}
    \tablefoot{(1) Galaxy name; (2) redshift; (3) metallicities derived from optical emission lines; (4) Lyman continuum escape fraction; and (5) spectral resolution of the observations. Dashes indicate that the quantities have not been measured.}
    \tablebib{
(a) \citet{pettini2004}; (b) \citet{izotov2016a}; (c) \citet{izotov2016b}; (d) \citet{leitherer2016}; (e) \citet{izotov2011}; (f) \citet{stark2008}; (g) \citet{borthakur2014}; (h) \citet{chisholm2017leak} (i) \citet{wuyts}}
    \end{table}

\section{Ultraviolet spectral fitting methods and results}
\label{sect:method}

We now describe the theory and methods we used to fit the UV spectra including stellar continua, ISM absorption lines, and Milky Way absorption lines. The method is then applied to simulated data (in particular to determine systematic errors) and to the observed spectra.

\subsection{Ultraviolet continuum and interstellar absorption line modeling}
\label{sect:metsimu}

\subsubsection{Adopted geometries and basic formulas}
\label{sect:geom}

To describe the radiation transfer in the host galaxy, we adopted the classical picket-fence model with different assumptions on the geometric distribution of gas and dust. In practice we considered two cases: {\em (a)} the picket-fence model with a uniform foreground dust screen, and {\em (b)} a clumpy picket-fence model where dust is only in the neutral gas clumps. Two parameters describe the two models: the dust attenuation (here parametrized by \ebv) and the geometric covering fraction of neutral gas ($C_f$), which is defined as the fraction of the total lines of sight of the emitted UV radiation that are intercepted by neutral gas in the direction toward the distant observer.

In {\em (a)} both the radiation emerging from the gas clumps (with geometric coverage $C_f$) and radiation directly escaping $(1-C_f)$ is attenuated by a uniform foreground dust screen. In {\em (b)}, only a fraction, $C_f$, of radiation is processed through the gas clumps, imprinting interstellar absorption lines and attenuating the stellar continuum. The rest escapes unaltered. Gas within these clumps is assumed to be homogeneous and the interclump medium has a negligible column density of neutral gas, i.e., is assumed to be completely transparent.

These simple models have already been examined and assumed by other authors \citep[e.g.,][]{zackrisson2013, borthakur2014,vasei2016,reddy2016stack}. We therefore only briefly list the main equations used in our spectral modeling. For a picket-fence model with a uniform foreground dust screen {\em (a)} the emergent flux $ F_\lambda$ is
    \begin{equation}
        F_\lambda = F^\star_\lambda \times 10^{-0.4 k_\lambda \ebv} \times \left(C_f \exp(-\tau_\lambda) + (1-C_f)\right),
        \label{eq:uniform}
    \end{equation}
where $ F^\star_\lambda$ is the intrinsic stellar emission prior to alteration by the ISM, $k_\lambda$ describes the attenuation law, and  $\tau_\lambda$ is the optical depth of the interstellar absorption lines. For a picket-fence model, with a clumpy gas distribution {\em (b)}, the emergent flux becomes
    \begin{equation}
        F_\lambda = F^\star_\lambda \times 10^{-0.4 k_\lambda \ebv} \times C_f \exp(-\tau_\lambda) + F^\star_\lambda \times (1-C_f),
        \label{eq:holes}
    \end{equation}    
where the second term describes the unattenuated, directly escaping radiation. This light is unattenuated because the holes are assumed to be free of gas and dust. For high covering fractions ($C_f \rightarrow 1$) or low attenuations (\ebv~$\rightarrow 0$), Eq.~\eqref{eq:uniform} and Eq.~\eqref{eq:holes} are identical.

We define the residual flux, $R$, as the ratio of the flux density at the observed wavelength of the line to the continuum flux density. The $R$ gives the fraction of light unabsorbed by the neutral gas. For saturated lines ($\tau_\lambda \gg 1$) the residual flux becomes
    \begin{equation}
       R=1-C_f
        \label{eq:d_uniform}
    \end{equation} 
for a uniform dust screen {\em (a)}, and
        \begin{equation}
       R=\frac{(1-C_f)}{10^{-0.4 k_\lambda \ebv} C_f + (1-C_f)}
        \label{eq:d_holes}
    \end{equation} 
for case {\em (b)} \citep[][]{vasei2016}. We note that R of the uniform screen model is always greater than, or equal to, the R of the clumpy model (see the discussion in Sect~\ref{dis:geometry}). 
     
Finally, the absolute escape fraction of radiation for monochromatic radiation, \fesc~$=F_\lambda/F^\star_\lambda$, becomes 
    \begin{equation}
       f_{\rm esc} = 10^{-0.4 k_{_\lambda} \ebv} \times \left(C_f\exp(-\tau_{_\lambda}) + (1-C_f)\right)
        \label{eq:f_uniform}
    \end{equation} 
    for the uniform dust screen {\em (a)}, and
    \begin{equation}
        f_{\rm esc} = 10^{-0.4 k_{_\lambda} \ebv} \times C_f \exp(-\tau_{_\lambda}) + (1-C_f) 
        \label{eq:f_holes}
    \end{equation} 
in clumpy geometry {\em (b)}. It is important to note that the values of $C_f$ and \ebv\ differ a priori between the two model geometries, except for the case of high covering fractions ($C_f \rightarrow 1$) or low attenuation (\ebv~$\rightarrow 0$). They must be determined consistently from fits to the observed data, adopting the equations corresponding to the assumed geometry. 
    
    \begin{table}
    \caption{Fitted absorption lines}
    \label{table:lines} 
    \centering                          
    \begin{tabular}{llllll}
    \hline \hline
       Ion & $\lambda_{\rm rest}$  [\AA] \\ \hline
       \multirow{8}{*}{\ion{H}{i}} & 920.947\tablefootmark{a}\\
        & 923.148\tablefootmark{a}   \\
        & 926.249 \tablefootmark{a}  \\
        & 930.751  \\
        & 937.814  \\
        & 949.742  \\
        & 972.517  \\
        & 1025.728   \\ \hline 
       \multirow{13}{*}{\ion{O}{i}} & 924.950\tablefootmark{a} \\
         &  929.517 \\
        &  930.257   \\
        &  936.629 \\
        &  948.686 \\
        &  950.885 \\
        &  971.738 \\
        & 976.448 \\
        & 988.578 \\ 
        & 988.655  \\
        & 988.773 \\
        & 1025.762 \\
        & 1039.230\\ 
        & 1302.168 \tablefootmark{b} \\\hline
         \multirow{2}{*}{\ion{O}{vi}} & 1031.912  \\
                 & 1037.613 \\ \hline
        \ion{C}{ii} &1036.337 \\ \hline
        \ion{C}{iii} & 977.030 \\ \hline
         \multirow{3}{*}{\ion{Si}{ii}} & 989.870 \\
         & 1020.70 \\
         & 1190.42\tablefootmark{c} \\
         & 1193.28\tablefootmark{c}\\
         & 1260.42\tablefootmark{c}\\ \hline
    \end{tabular}  
    
    \tablefoot{Wavelengths are in vacuum.\\ \tablefoottext{a}{Used only for generating synthetic spectra, see Sect. \ref{sect:data}}.\\
    \tablefoottext{b}{Used for \ion{O}{i} measurements in Tol0440-381 and Mrk54 spectra, see Sect. \ref{sect:oi}}.\\
    \tablefoottext{c}{Used only to measure \ion{Si}{ii} covering fraction, see Sect.\ref{method:si2}}.}
    \end{table}
    
\subsubsection{Fitting method}  
 \label{fitting}
\begin{figure}
   \centering
   \includegraphics[width=\hsize]{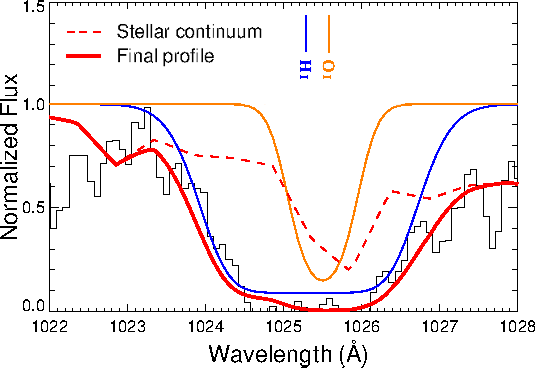}
      \caption{Total fit (red solid line) of the Ly$\beta$ absorption line for the galaxy GP~1244+0216 (black data). The contributions from the stellar continuum (dashed red line), Ly$\beta$ absorption line (blue solid line), and \ion{O}{i} absorption line (orange solid line) blend together. \ion{O}{i} absorption lines at 989\AA\ and 1039\AA\ are unblended and robustly constrain the \ion{O}{i} profile. All three components need to be accounted for when fitting the Lyman series.}
         \label{fig:lybzoom}
   \end{figure}

The main spectral region modeled here is the Lyman series, \ion{H}{i} absorption lines from Ly$\beta$ to the Lyman break ($\approx 912-1050\AA$). In practice, owing to the lower S/N close to the Lyman break, the bluest Lyman line that we include is Ly6 (930~\AA). Figure~\ref{fig:lybzoom} emphasizes the complicated nature of the reddest Lyman series line, Ly$\beta$: strong stellar continuum features (red dashed line) blend with the broad \ion{H}{i} interstellar absorption (blue line) and the weaker \ion{O}{i} interstellar absorption line (orange line). Consequently, bluer Lyman series lines (especially Ly$\gamma$) have fewer complications because of their simpler stellar continua.   

We modeled ISM metal absorption lines from \ion{O}{i}, \ion{O}{vi}, \ion{Si}{ii}, \ion{C}{ii} and  \ion{C}{iii} (listed in Table~\ref{table:lines}). \ion{O}{i} has a similar ionization structure as \ion{H}{i}, such that \ion{O}{i} absorption lines blend with all \ion{H}{i} lines, except for \ion{O}{i}~989 and 1039~$\angstrom$. These two lines constrain the \ion{O}{i} profile.

First, we fit the stellar continuum. We started with an initial stellar model using a linear combination of ten single-age stellar continuum models with ages of 1, 2, 3, 4, 5, 8, 10, 15, 20, and 40 Myr. We also used stellar continuum metallicities of 0.05, 0.2, 0.4, 1, or 2 Z$_\odot$. These spectra were drawn from the fully theoretical {\small STARBURST99} library \citep[S99;][]{leitherer1999}, computed with the WM-Basic method \citep{leitherer2010},  and have a spectral resolution R(S99) $\approx$ 2500. We chose the stellar continuum metallicity closest to the ISM metallicity (Table~\ref{table:sample}). The {\small STARBUST99} models use a Kroupa initial mass function with a high (low) mass exponent of 2.3 (1.3), a high-mass cutoff at 100~M$_\odot$, and the stellar evolution tracks with high mass loss from \citet{meynet1994}.  We fit for a linear combination of the stellar continuum flux, $F_{S99}$, as
\begin{equation}
     F^\star = \Sigma_{i=1}^{10}{X_i \, F^{99}_i} ,
     \label{eq:stellarflux}
\end{equation}
\noindent where $X_i$ are the linear coefficients for a given age ($i$) and $F^{99}_i$ are the {\small STARBURST99} theoretical stellar continuum models for a given age.

Absorption lines of different ions are added using Voigt profiles defined by the velocity shift $v$, $b$-parameter, column density $N$, and $C_{f}$. The metal covering fraction is initially fixed to 1 to reduce the number of free parameters.  We included the \ion{O}{i} absorption lines that are directly blended with the Lyman series. Each ion is considered independent, as are its parameters. For each galaxy, we tested whether including the remaining metal lines listed in Table~\ref{table:lines} improves the Lyman series fits. If, by eye, they did not improve the fits, we did not include the lines. Finally, the {\small STARBURST99} models are convolved with the nominal spectral resolution of the observations.

We accounted for dust attenuation using the attenuation law from \citet{reddy2016dustlaw}, a uniform dust screen model, and fitting for the dust attenuation parameter \citep[\ebv; similar to][]{chisholm2015}. The linear combination of stellar continuum models, interstellar absorption lines, and dust attenuation produces the final fitted spectrum.

We fit the data using an IDL routine based on the nonlinear least squares method, {\small MPFIT} \citep{markwardt2009}. The {\small MPFIT} method returns the best fit and errors for \ebv, $b$, $v$, $N,$ and $C_f$ of each ion. For observed spectra, the first step consists of masking the ISM absorption lines and the contaminating Milky Way absorption lines, geocoronal emission, and intervening absorbers. We applied these masks on the data and fit for the linear combination of dust-attenuated {\small STARBURST99} models. In a second step, we fixed the stellar continuum and fit for the ISM absorption lines and Milky Way absorption lines adjacent to the ISM lines. We fit for all of the observed Lyman series lines up to Ly6, provided that they are not near geocoronal emission or intervening absorbers, and do not have a S/N below one. Since the simulated data do not contain the extra complications of adjacent Milky Way and geocoronal lines, all parameters are simultaneously fit in one step.

\subsection{Fitting simulated data}
\label{meth:simu}

We tested our fitting method with both noise-free and noisy synthetic data to determine how well the estimated parameter errors represent the actual errors. Additionally, we tested how these errors depend on the S/N and spectral resolution. In Appendix~\ref{sect:data} we fully describe the generation of synthetic spectra, but here we summarize the steps. The synthetic spectra were produced for two scenarios: one assuming the picket-fence model ($\log(\nh$[cm$^{-2}$]) = 20, $C_f = 0.9$) and one describing a uniform ISM in the density-bounded regime ($\log(\nh[$cm$^{-2}]$) = 17.57, $C_f = 1$). Both scenarios correspond to an escape fraction $\fesc=0.1$, regardless of the chosen dust distribution, since we set \ebv~$= 0 $ (Eqs.~\eqref{eq:f_uniform} and \eqref{eq:f_holes}).

We created synthetic spectra using the parameters of the picket-fence and density-bounded regime (see Table~\ref{table:species}) for 7 different spectral resolutions between $R = 600-15000$. For each of these 14 set-ups (7 spectral resolutions and two scenarios) we created 50 different realizations by adding  various sets of random Gaussian noise. The level of Gaussian noise was chosen to produce a final S/N per pixel between $2-50$, in 7 total S/N steps. In total we created 98 different configurations (2 different scenarios, 7 spectral resolutions, and 7 S/N ratios), and each configuration has 50 individual spectra, for a total of 4900 synthetic spectra.  Figs.~\ref{fig:singlefit15000} and~\ref{fig:singlefit600} show synthetic spectra and the best-fit models for high and low spectral resolution ($R = 15000$ and $R = 600$) for three S/N configurations (noise-free, 10 and 2).

We fit the synthetic spectra with the methods outlined in Sect.~\ref{fitting}. From this fitting we define two types of error: 
\begin{itemize}
     \item The \textit{statistical error} ($Err_{\rm stat}$). This is the individual errors returned by {\small MPFIT}. 
    \item The \textit{systematic error} ($Err_{\rm syst}$). This is the deviation of the 50 parameter estimates of each scenario from the actual parameter value that created the original line profile.\ The $Err_{\rm syst}$ is therefore a function of S/N and resolution. $Err_{\rm syst}$ is calculated as 
    \begin{equation}
        Err_{\rm syst} = \sqrt{\frac{1}{50}\sum_{i=1}^{50}{(x_i - x_{T})^2}},
    \end{equation}
     where $x_i$ is the estimated parameter and $x_{T}$ is the true parameter value. The systematic error accounts for differences between the estimated parameters and the actual parameters that are unaccounted for by {\small MPFIT}. We express $Err_{\rm syst}$ as a percent error, defined as the deviation of the measured parameter from the true parameter divided by the true parameter value. This fractional error is more broadly applicable to a variety of measurement values.
\end{itemize}
By definition, $Err_{\rm syst}$ is not included in the statistical uncertainties reported by {\small MPFIT}, but it critically describes the ability of {\small MPFIT} to recover the actual parameter values. We account for $Err_{\rm syst}$ by defining the total error of each parameter ($Err_{tot}$) as the quadratic sum of $Err_{\rm syst}$ and $Err_{\rm stat}$ as follows:
\begin{equation}
        Err_{\rm tot} = \sqrt{Err_{\rm syst}^2 + Err_{\rm stat}^2}.
        \label{eq:totalerr}
    \end{equation}
The systematic errors derived here are included in the errors of the \ion{H}{i} covering fraction and column density (Cols.~3 and 6 of Table~\ref{table:hiresult})

\subsection{Fitting observed data}
\label{method:obs}

Using the fitting method described in Sect.\ \ref{fitting}, we fit the observed spectra of the 18 galaxies in our sample. The resulting fit parameters \ebv, $C_f$, $v$, $b$, and $N$ are listed in Table \ref{table:hiresult}, and the corresponding best-fit spectra are shown in Appendix~\ref{app:fits}. The reported uncertainties on the \ion{H}{i} column densities and covering fractions account for the systematic errors derived from simulations.

We note that Tol0440-381 and Mrk54 have much lower redshifts than the rest of the sample. Consequently there are more Milky Way absorption lines near the Lyman series. This high density of Milky Way Lyman series absorption lines means that we cannot accurately fit the Lyman series with Voigt profiles. Rather, we used nonparametric fits to describe the \ion{H}{i} properties (see Sect.~\ref{method:depth}).

The \ion{H}{i} column densities are affected by large uncertainties because the low S/N and insufficient resolution of the spectra do not allow a reliable estimate of \nh\ given the saturation of the \ion{H}{i} absorption lines (see Sect.~\ref{simcold}). Therefore direct \nh\ measurements from the Lyman series are not reliable and an indirect method is used (Sect.\ \ref{sect:oi}). To examine whether saturation and  resulting degeneracies affect the $C_f$ or metal column densities that we derived, we refit the spectra fixing the \ion{H}{i} column density to $\nh = 10^{18}$~cm$^{-2}$. This \nh\ corresponds to an optically thick portion of the curve of growth for the observed Lyman series lines (from Ly$\beta$ to Ly6). We find that all the fit parameters (except for $b$(\ion{H}{i})) are consistent within $1 \sigma$ with the results listed in  \autoref{table:hiresult}. The fact that $b$ changes is consistent with $b$ being degenerate with \nh\ at these column densities. This indicates that saturated \ion{H}{i} absorption does not affect the measured velocities and covering fractions.

    \begin{table*}
    \caption{Derived \ion{H}{i} properties from the Lyman series absorption lines.}
    \label{table:hiresult}
    \centering   
     \begin{tabular}{lllllllllll}
    \hline \hline
    Galaxy name & \ebv\  & $\log(\nh)$ & $b$ & $v$ & $C_f$ fits & $C_f$ depth &  $C_f$ final \\
     & & [log(cm$^{-2}$)] & [km s$^{-1}$] &[km s$^{-1}$] &  \\
     (1) & (2) & (3) & (4) & (5) & (6) & (7) & (8) \\ \hline 
    J0921+4509  & 0.222 $\pm$ 0.015 & 17.94 $\pm$ 3.22 & 95 $\pm$ 49 &   -56 $\pm$ 13 & 0.754 $\pm$ 0.111 & 0.769 $\pm$ 0.116 & 0.761 $\pm$ 0.080\\ 
    J1503+3644  & 0.274 $\pm$ 0.014 & 20.50 $\pm$ 2.33 &  91 $\pm$ 24 &  -127 $\pm$ 12 & 0.634 $\pm$ 0.109 & 0.754 $\pm$ 0.062 & 0.725 $\pm$ 0.054\\ 
    J0925+1409  & 0.164 $\pm$ 0.015 & 17.81 $\pm$ 3.26 &  81 $\pm$ 73 &  -214 $\pm$ 151& 0.652 $\pm$ 0.218 & 0.635 $\pm$ 0.094 & 0.638 $\pm$ 0.086\\ 
    J1152+3400  & 0.134 $\pm$ 0.022 & 20.94 $\pm$ 2.70 & 187 $\pm$ 48 &  -468 $\pm$ 39 & 0.506 $\pm$ 0.067 & 0.619 $\pm$ 0.08 & 0.548 $\pm$ 0.053\\ 
    J1333+6246  & 0.151 $\pm$ 0.043 & 21.14 $\pm$ 1.46 & 102 $\pm$ 19 &  -126 $\pm$ 48 & 0.731 $\pm$ 0.122 & 0.826 $\pm$ 0.066 & 0.804 $\pm$ 0.058\\ 
    J1442-0209  & 0.140 $\pm$ 0.015 & 19.60 $\pm$ 3.53 & 123 $\pm$ 183 &  -261 $\pm$ 34 & 0.621 $\pm$ 0.120 & 0.549 $\pm$ 0.040 & 0.556 $\pm$ 0.038\\ 
    Tol1247-232 & 0.156 $\pm$ 0.010 & 21.89 $\pm$ 3.09 & 66 $\pm$ 87 &  260 $\pm$ 31 & 0.587 $\pm$ 0.061 & 0.690 $\pm$ 0.084 & 0.623 $\pm$ 0.049\\ 
    Tol0440-381 & 0.271 $\pm$ 0.028 & -                & -            &  -            &  -                & 0.570 $\pm$ 0.084 & 0.570 $\pm$ 0.084\\ 
    Mrk54       & 0.359 $\pm$ 0.013 &  -                & -            &  -            &  -              & 0.504 $\pm$ 0.077  & 0.504 $\pm$ 0.077\\ 
    J0926+4427  & 0.114 $\pm$ 0.010 & 16.39 $\pm$ 0.40 & 213 $\pm$ 23 &  -199 $\pm$ 12 & 0.723 $\pm$ 0.048 & 0.814 $\pm$ 0.048 & 0.768 $\pm$ 0.034\\ 
    J1429+0643  & 0.108 $\pm$ 0.015 & 16.17 $\pm$ 0.50 & 236 $\pm$ 56 &  -241 $\pm$ 36 & 0.897 $\pm$ 0.071 & 0.960 $\pm$ 0.061 & 0.930 $\pm$ 0.046\\ 
    GP0303-0759 & 0.121 $\pm$ 0.045 & 16.07 $\pm$ 1.50 & 192 $\pm$ 99 &  -266 $\pm$92& 0.908 $\pm$ 0.207 & -  & 0.908 $\pm$ 0.207\\ 
    GP1244+0216 & 0.290 $\pm$ 0.043 & 16.34 $\pm$ 0.80 & 220 $\pm$ 79 &  -78 $\pm$ 48  & 0.909 $\pm$ 0.357 & 0.950 $\pm$ 0.131 & 0.946 $\pm$ 0.123\\ 
    GP1054+5238 & 0.204 $\pm$ 0.044 & 19.60 $\pm$ 3.81 & 164 $\pm$ 78 &  -166 $\pm$ 29 & 0.702 $\pm$ 0.131 & 0.889 $\pm$ 0.158 & 0.778 $\pm$ 0.101\\ 
    GP0911+1831 & 0.352 $\pm$ 0.038 & 16.76 $\pm$ 1.19 & 188 $\pm$ 43 &  -273 $\pm$ 40 & 0.731 $\pm$ 0.150 & 0.765 $\pm$ 0.116 & 0.752 $\pm$ 0.092\\
    \sone\      & 0.201 $\pm$ 0.001 & 17.16 $\pm$ 1.06 & 380 $\pm$ 24 &  -264 $\pm$ 21 & 0.932 $\pm$ 0.038 & 0.998 $\pm$ 0.009 & 0.994 $\pm$ 0.009\\   
    \stwo\      & 0.370 $\pm$ 0.002 & 17.15 $\pm$ 1.22 & 269 $\pm$ 47 &  -208 $\pm$ 25 & 1.000 $\pm$ 0.075 & 0.990 $\pm$ 0.038 & 0.992 $\pm$ 0.034\\ 
    Cosmic Eye  & 0.405 $\pm$ 0.001 & 22.59 $\pm$ 1.28 & 199 $\pm$ 5  &    56 $\pm$ 16 & 0.918 $\pm$ 0.072 & 0.998 $\pm$ 0.024 & 0.990 $\pm$ 0.023\\ \hline
    \end{tabular}
    \tablefoot{(1) Galaxy name; (2) dust attenuation parameter (\ebv); (3) logarithm of the \ion{H}{i} column density; (4) \ion{H}{i} Doppler $b$-parameter; (5) \ion{H}{i} velocity shift; and (6) \ion{H}{i} covering fraction from the fits. Uncertainties of the \ion{H}{i} column density and covering fraction from the fits include the systematic errors detailed in Sect~\ref{simcf}. (7) \ion{H}{i} covering fraction measurements derived from the residual flux of the individual \ion{H}{i} absorption lines (Sect.~\ref{method:depth}; Table~\ref{table:cfdepth}). GP0303-0759 does not have a reliable depth measurement because Milky Way absorption lines overlap the Lyman series; (8) weighted mean of (6) and (7).}
    \end{table*}

\subsubsection{Nonparametric measurements of the \ion{H}{i} covering fraction}
\label{method:depth}

Fitting the Lyman series using the the radiative transfer equation (Eq.\ \eqref{eq:uniform}) is one way to measure the \ion{H}{i} covering fraction. However, this assumes that the absorption profiles follow a single Lorentzian velocity distribution. Observed absorption profiles arising from galactic outflows are highly non-Lorentzian \citep[][]{heckman2000, pettini02, shapley2003, weiner09, chisholm2017mass}. Consequently, we also measured $C_f$ from the residual flux of the Lyman series lines after removing the stellar continuum (see Eq.~\eqref{eq:d_uniform}). This assumes a uniform dust geometry and that the Lyman series is fully saturated. The Lyman series is saturated from $\nh \ga 10^{16} $~cm$^{-2}$ (for Ly$\beta$ to Ly6). Importantly, this nonparametric approach does not assume a velocity distribution of the \ion{H}{i} gas and accounts for arbitrary line profiles.  In other words, the nonparametric approach does not assume that the gas has the same covering fraction at every velocity \citep[cf. discussion in][]{vasei2016}.

The covering fraction is derived as the maximum of $(1-F_{\rm Gau})$ in a velocity range chosen by eye near the deepest part of each Lyman series line. The $F_{\rm Gau}$ is the stellar continuum removed flux, modified by a Gaussian kernel centered on zero with standard deviation corresponding to the error array. We measured $(1-F_{\rm Gau})$ 1000 times,  where each time the flux value of each pixel is determined from a different noise distribution. We then took the median and standard deviation of this distribution as the $C_f$ value and uncertainty for each Lyman series transition. Table~\ref{table:cfdepth} lists the $C_f$ derived from the residual flux of each Lyman series transition in each galaxy. We then defined the $C_f$(\ion{H}{i}) from the residual flux as the error weighted mean ($M_W$) of the $i$ observed Lyman series transitions and the $C_f$ error as the error on $M_W$ as
\begin{equation}
\begin{aligned}
    M_W &= \frac{\sum_{i=1}^{n} {C_f}_{i} \times \omega_i}{\sum_{i=1}^{n} \omega_i} \; \mbox{with} \; \omega_i = \frac{1}{\sigma_i^2}, \\
    \sigma_{M_W} &= \sqrt{\frac{1}{\sum_{i=1}^{n} \omega_i}}.
    \label{eq:wmean}
\end{aligned}
\end{equation}
The corresponding values, denoted as $C_f$ depth, are reported in Table~\ref{table:hiresult}. The $C_f$ depth values are consistent, within the errors, with the $C_f$ values derived from the fits in Sect.~\ref{method:obs}. Except for Tol0440-381 and Mrk54, where we did not fit the absorption lines with the method in Sect.~\ref{fitting}, we computed the final $C_f$(\ion{H}{i}) as the weighted mean of the $C_f$ depth and the fitted $C_f$ values (column 8 in Table~\ref{table:hiresult}).

\subsubsection{Indirect measurements of the \ion{H}{i} column density}
\label{sect:oi}
Since the Lyman series lines are saturated, but not damped,  and the spectra have insufficient spectral resolution and S/N direct \ion{H}{i} column density measurements are largely unconstrained, as discussed further in Sect.~\ref{simcold}.
Therefore, indirect methods of measuring \nh\ are needed. We used the \ion{O}{i} absorption lines, constrained when possible by the unsaturated \ion{O}{i}~1039\AA\ absorption line, to derive the \ion{O}{i} column density.   Using the known metallicity (\oh) of the galaxy, we then indirectly inferred the hydrogen column density. This approach assumes that the emission-line based oxygen abundance, tracing the chemical composition of the ionized gas, is identical to that of the neutral gas.
 If the metallicity was lower in the neutral gas, for example,\ because of the presence of some pristine gas \citep[see, e.g.,][and references therein]{Lebouteiller} or incomplete mixing of the metals (see Sect.~\ref{sect:cf}), the resulting \nh\ would be higher (and more saturated) than inferred here.

The \ion{O}{i} fit parameters $b$, $v$, $N_{\ion{O}{i}}$, and the  \ion{H}{i} column density derived using the metallicities from Table~\ref{table:sample}, are listed in  Table~\ref{table:oiresult}. A curve of growth analysis indicates that \ion{O}{i}~1039\AA\ saturates at $N_{\ion{O}{i}}$~>~$10^{16.5}$~cm$^{-2}$, whereas all but one of our $N_{\ion{O}{i}}$ are less than $10^{16.2}$~cm$^{-2}$. If the \ion{O}{i}~1039 line is saturated, then the \nh\ values in Table~\ref{table:oiresult} would underestimate the actual \nh\ (i.e., \nh\ is not lower than the quoted values).   Fig.~\ref{fig:J0921} demonstrates that \ion{O}{i}~1039\AA\ is resolved from the \ion{C}{ii}~1036\AA\ and the \ion{O}{vi}~1038\AA\  absorption lines. However, \ion{O}{i} is undetected for J0925+1409 (see Fig.~\ref{fig:J0925}), and we used \ion{O}{i} 1302~\AA\ for Tol0440-381 and Mrk54. We could not determine \nh\ for the three \megasaura\ sources, since the \ion{O}{i} line was either saturated, or there was not a literature metallicity. The resulting \ion{H}{i} column densities are found to be $\log(\nh) \sim 18.6-20$ cm$^{-2}$.

We fixed the \ion{O}{i} covering fractions to the \ion{H}{i} values (i.e., $C_f(\text{\ion{O}{i}}) = C_f(\text{\ion{H}{i}})$). This is plausible because \ion{O}{i} and \ion{H}{i} have similar ionization potentials and their ionization fractions are locked together by charge exchange. We examined how tying the covering fractions impact the derived \ion{O}{i} column densities. As an extreme case, we fixed the \ion{O}{i} covering fraction to 1 for J1152+3400, which has one of the lowest $C_f$(\ion{H}{i}), and J0921+4509, which has the lowest \nh\ derived from \ion{O}{i}. We found that $\log(N_\text{OI})$ is reduced by 0.30 and 0.15 dex, respectively, which is comparable to the $N_{\ion{O}{i}}$ errors. Therefore, we conclude that tying $C_f$(\ion{O}{i}) to  $C_f$(\ion{H}{i}) does not drastically change the measured \ion{O}{i} column density.

    \begin{table*}
    \caption{Fitted \ion{O}{i} properties}
    \label{table:oiresult}
    \centering   
    \begin{tabular}{lllll}
    \hline \hline
    Galaxy name&   $\log(N_{\ion{O}{i}})$ & $b$ & $v$ & $\log(\nh)$ \\
      & [log(cm$^{-2}$)] & [km s$^{-1}$] & [km s$^{-1}$]  & [log(cm$^{-2}$)] \\
      (1) & (2)& (3)& (4) & (5)  \\     \hline
     J0921+4509  &  15.30 $\pm$ 0.13     &  45 $\pm$ 15  &  62  $\pm$ 11   & 18.63  $\pm$ 0.19 \\ 
     J1503+3644  &  15.55 $\pm$ 0.16     & 302 $\pm$ 98  &  102  $\pm$ 77  & 19.60  $\pm$ 0.17 \\ 
     J0925+1409  &  -                    & -             &  -              & -                 \\ 
     J1152+3400  & 15.43 $\pm$ 0.17      & 227 $\pm$ 129 & -102  $\pm$ 83 & 19.43  $\pm$ 0.18 \\ 
     J1333+6246  & 15.54 $\pm$ 0.35      & 287 $\pm$ 277 & -213 $\pm$ 152 & 19.78  $\pm$ 0.37 \\ 
     J1442-0209  & 15.62 $\pm$ 0.58      & 178 $\pm$ 145 & 82 $\pm$ 101     & 19.69  $\pm$ 0.58 \\ 
     Tol1247-232 & 15.29 $\pm$ 0.43      & 278 $\pm$ 168  & 58 $\pm$ 143  & 19.19  $\pm$ 0.44\\ 
     Tol0440-381\tablefootmark{a} & 15.47 $\pm$ 0.02    & 623 $\pm$ 28  & 12 $\pm$ 19  & 19.27  $\pm$ 0.10 \\ 
     Mrk54\tablefootmark{a}       & 16.07 $\pm$ 0.01     & 619 $\pm$ 8 & 15 $\pm$ 6     & 19.37  $\pm$ 0.10 \\
     J0926+4427  & 15.77 $\pm$ 0.02      & 118 $\pm$ 5   &  -141 $\pm$ 4   & 19.76  $\pm$ 0.05 \\ 
     J1429+0643  & 15.55 $\pm$ 0.24      & 218 $\pm$ 74  &  -43 $\pm$ 98  & 19.35  $\pm$ 0.25 \\ 
     GP0303-0759 & 15.41 $\pm$ 0.18      & 209 $\pm$ 84  &   31 $\pm$ 70   & 19.55  $\pm$ 0.19 \\  
     GP1244+0216 & 16.12 $\pm$ 0.14      & 157 $\pm$ 47  &  -67 $\pm$ 36   & 19.95  $\pm$ 0.15 \\ 
     GP1054+5238 & 15.73 $\pm$ 0.14    & 256 $\pm$ 72  &     39 $\pm$ 55   & 19.63  $\pm$ 0.15 \\ 
     GP0911+1831 & 15.73 $\pm$ 0.15      & 221 $\pm$ 79  &  -102 $\pm$  67  & 19.73  $\pm$ 0.16 \\
    \sone        & 16.11 $\pm$ 0.03     & 200 $\pm$ 7  & -290 $\pm$ 11   & -                 \\
    \stwo        & 15.57 $\pm$ 0.13      & 133 $\pm$ 44  & -89 $\pm$ 33     & -                 \\    
    Cosmic Eye   & 20.64\tablefootmark{b} $\pm$  0.66 & 24  $\pm$ 77   & 208 $\pm$ 9   & 24.04\tablefootmark{b}  $\pm$ 0.67 \\ \hline
    \end{tabular}
    \tablefoot{(1) Galaxy name; (2) logarithm of \ion{O}{i} column density; (3) \ion{O}{i} Doppler b-parameter; (4) velocity shift of the \ion{O}{i} line. The \ion{O}{i} covering fraction is fixed to the \ion{H}{i} final value from Table~\ref{table:hiresult}. We did not detect the \ion{O}{i} absorption lines in the J0925+1409 spectra. (5) \ion{H}{i} column density derived from the product of (3) and \oh\ (Table~\ref{table:sample}). We cannot estimate \nh\ for \sone\ and \stwo\ because they do not have a measured \oh.  \\
    \tablefoottext{a}{Used \ion{O}{i} 1302 \AA} \\
    \tablefoottext{b}{Saturated \ion{O}{i} line, hence unreliable \nh\ determination. From the damped \lya\ profile, \cite{quider2010} obtained $\nh=(3.0 \pm 0.8) \times 10^{21}$ cm$^{-2}$.}}
    \end{table*}

\subsubsection{The \ion{Si}{ii} covering fraction}
\label{method:si2}

We measured the \ion{Si}{ii} covering fraction ($C_{f}$(\ion{Si}{ii})) using the \ion{Si}{ii}~1190~$\angstrom$ doublet and the \ion{Si}{ii}~1260~$\angstrom$ singlet. For the \ion{Si}{ii}~1260~$\angstrom$ singlet, $C_{f}$(\ion{Si}{ii}) was derived using the same procedure as for the residual flux measurements of the Lyman series (Sect.~\ref{method:depth}).  A different approach, detailed in \citet{chisholm2017mass}, was adapted for the doublet. As the two transitions share the same $C_f$ and the ratio of the two optical depths is given by the ratio of their oscillator strengths, $f$, the velocity-resolved covering fraction was measured from a system of equations \citep{hamann1997} as 
\begin{equation}
    C_f(v) = \frac{F_W(v)^2 - 2F_W(v) +1}{F_S(v) - 2 F_W(v) +1}
    \label{eq:fcovvel}
,\end{equation}
\noindent where $F_W$ is the continuum subtracted flux of the weaker doublet line (\ion{Si}{ii} 1190~$\angstrom$) and $F_S$ is the continuum subtracted flux of the stronger doublet line (\ion{Si}{ii} 1193~$\angstrom$). This method accounts for the possibility that the lines are not saturated. We measured the covering fraction 1000 times by varying the flux in a similar way to the residual flux measurements of the Lyman series. The median and standard deviation of this distribution is taken as the $C_f$(\ion{Si}{ii}) value and error (Table~\ref{table:sicf}). Using the \ion{Si}{ii} 1260 \AA\ singlet assumes that the line is saturated, whereas the doublet method accounts for unsaturated absorption. We derived statistically consistent values using both methods (all within 1$\sigma$) for all the galaxies for which we can measure $C_f$(\ion{Si}{ii}) with both methods. This indicates that \ion{Si}{ii} 1260~\AA\ is indeed saturated.

    \begin{table*}
    \caption{\ion{Si}{ii} covering fractions}
     \label{table:sicf}
    \centering   
    \begin{tabular}{llllllll}
    \hline \hline
    Galaxy name&   \ion{Si}{ii} 1190~\AA & \ion{Si}{ii} 1260~\AA  & Mean \\ 
    (1) & (2) & (3) &(4) \\ \hline 
    J0921+4509     & 0.66 $\pm$ 0.32 & 0.59 $\pm$ 0.15 & 0.60 $\pm$ 0.14  \\ 
    J1503+3644     & 0.38 $\pm$ 0.35 & 0.56 $\pm$ 0.45 & 0.45 $\pm$ 0.28   \\
    J0925+1409     & 0.40 $\pm$ 0.28 & 0.43 $\pm$ 0.34 & 0.41 $\pm$ 0.19  \\ 
    J1152+3400     & 0.27 $\pm$ 0.26 & 0.23 $\pm$ 0.55 & 0.27 $\pm$ 0.24   \\ 
    J1333+6246     & 0.29 $\pm$ 0.26 & 0.56 $\pm$ 0.35 & 0.39 $\pm$ 0.21   \\
    J1442-0209     & 0.46 $\pm$ 0.23 & 0.48 $\pm$ 0.34 & 0.47 $\pm$ 0.19  \\
    Tol1247-232    &  - & 0.26 $\pm$ 0.01 & 0.26 $\pm$ 0.01             \\ 
    Tol0440-381    &  -  & 0.37 $\pm$ 0.04 & 0.37 $\pm$ 0.04               \\ 
    Mrk54          &  -  & 0.32 $\pm$ 0.01 & 0.32 $\pm$ 0.01              \\ 
    J0926+4427     & 0.37 $\pm$ 0.19 & 0.36 $\pm$ 0.06 & 0.36 $\pm$ 0.06 \\ 
    J1429+0643     & 0.73 $\pm$ 0.26 & 0.78 $\pm$ 0.11 & 0.77 $\pm$ 0.10 \\ 
    GP0303-0759    & 0.54 $\pm$ 0.29 & 0.47 $\pm$ 0.12 & 0.48 $\pm$ 0.11\\ 
    GP1244+0216    & 0.49 $\pm$ 0.39 & 0.51 $\pm$ 0.18 & 0.50 $\pm$ 0.16 \\ 
    GP1054+5238    & 0.41 $\pm$ 0.34 & 0.49 $\pm$ 0.22 & 0.46 $\pm$ 0.19 \\ 
    GP0911+1831    & 0.39 $\pm$ 0.36 & 0.45 $\pm$ 0.33 & 0.43 $\pm$ 0.24 \\
    \sone\         & 0.97 $\pm$ 0.06 & 0.91 $\pm$ 0.04 & 0.93 $\pm$ 0.03  \\ 
    \stwo\         & 0.83 $\pm$ 0.16 & 0.79 $\pm$ 0.10 & 0.80 $\pm$ 0.08 \\ 
    Cosmic Eye     & 0.98 $\pm$ 0.06 & 0.94 $\pm$ 0.03 & 0.95 $\pm$ 0.02 \\ \hline
    \end{tabular}
    \tablefoot{(1) Galaxy name; (2) $C_f$(\ion{Si}{ii}) derived from the \ion{Si}{ii}~1190~$\angstrom$ doublet using Eq.~\eqref{eq:fcovvel}; (3) $C_f$(\ion{Si}{ii}) derived from the \ion{Si}{ii}~1260~$\angstrom$ absorption line; (4) weighted mean between (2) and (3). We do not observe the \ion{Si}{ii}~1190~$\angstrom$ doublet for Tol1247-232, Tol0440-381, and Mrk54 because these lines fall in the COS detector gap.}
    \end{table*}

\subsection{Effect of  ultraviolet attenuation law on covering fraction}

A priori the measurements made here also depend on the dust attenuation law used. In this study, we used the attenuation law from \citet{reddy2016dustlaw} because it is defined blueward of Ly$\alpha$. We refit J1503+3644 and GP0911+1831, which are two galaxies with high \ebv\, and low $C_f$, using a Small Magellanic Cloud (SMC) attenuation law\footnote{Values have been taken from the IDL routine from J. Xavier Prochaska: \url{https://github.com/profxj/xidl/tree/master/Dust}} still assuming a uniform dust foreground. The SMC law  is significantly steeper than the \citet{reddy2016dustlaw} law. With the SMC dust law, we measured $C_f$(\ion{H}{i}) of 0.653 $\pm$ 0.109 and 0.744 $\pm$ 0.156, respectively. These are consistent, within 1 $\sigma$, with the $C_f$(\ion{H}{i}) estimated using the \citet{reddy2016dustlaw} attenuation law (0.634 $\pm$ 0.109 and 0.731 $\pm$ 0.150). The \ebv\ values change based upon the attenuation law used to match the observed continuum,  but these changes do not affect the measured $C_f$. We therefore conclude that the adopted attenuation law does not significantly impact the measured covering fractions.

\section{Recovering \ion{H}{i} properties from simulated spectra}
\label{sect:sims}

We simulated synthetic spectra and fit these mock \ion{H}{i} lines with the method in Sect.~\ref{meth:simu}. Comparing the fitted results with the parameters that created the spectra characterizes how accurately the method returns the \ion{H}{i} parameters. We discuss these results in the context of the \ion{H}{i} column density (Sect.~\ref{simcold}) and the covering fraction (Sect.~\ref{simcf}). This discussion illustrates that $C_f$ is accurately measured for most resolutions and S/N ratios, while the \ion{H}{i} column density has large uncertainties. These simulations are especially helpful for planning future observations by determining the S/N ratios and resolutions required to accurately measure the covering fractions of LyC emitters. 

\subsection{\ion{H}{i} column densities}
\label{simcold}
The synthetic spectra allow us to quantify how accurately our method reproduces the \ion{H}{i} properties. For S/N ratios (per pixel) less than 10 and resolutions less than 3000, the simulations have \nh\ percent errors greater than 300\%  (Table~\ref{table:density}). Even at higher resolutions ($R=15000$), the percent error is greater than 200\%, unless the S/N is greater than 5. For the lower S/N ratios typical of our observations, we measured order of magnitude systematic uncertainties on \nh.  These large uncertainties are inherent because the Lyman series transitions saturate for \ion{H}{i} column densities between $N = 10^{16}$~cm$^{-2}$ to $N = 10^{22} $cm$^{-2}$. 
For these so-called Lyman limit systems, high quality and very high-resolution spectra ($R \sim 30000$) are needed to constrain \nh\ with Voigt fitting methods \citep[see, e.g.,][]{omeara2007}. Therefore, we conclude that \nh\ cannot be directly fitted from the Lyman series absorption lines.
However, the \ion{O}{i} absorption lines included in our fits remain unsaturated for $N_{\ion{O}{i}} <10^{16.5}$~cm$^{-2}$ and do not suffer from these large uncertainties. Consequently, the neutral column density is most accurately inferred by converting $N_{\ion{O}{i}}$ into \nh\ using the gas-phase metallicity,
as in Sect.~\ref{sect:oi}.

\subsection{\ion{H}{i} covering fractions}
\label{simcf}

\begin{figure}
\centering 
\includegraphics[width=\linewidth]{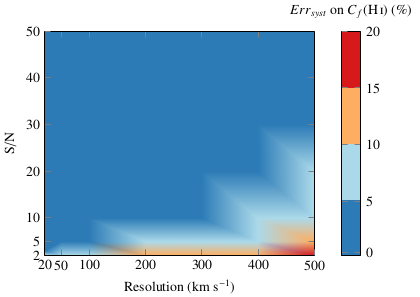}
\caption{Color map of the systematic percent error of the covering fraction as a function of the resolution and S/N. The synthetic spectra are created with R < 120 km s$^{-1}$ combine a theoretical stellar continuum spectra with R(S99) = 120 km s$^{-1}$ and absorption lines of spectral resolution R. The covering fraction is recovered to within 5\% of the estimated parameter for all observations within the dark blue region. }
\label{figure:picket}
\end{figure}

Conversely, the simulations show that $C_f$(\ion{H}{i}) has a low percent error, under typical observing conditions. At $R>1500$ and S/N~$> 5$, the $C_f$ systematic percent errors are less than 6\% of the measured value (Fig.~\ref{figure:picket}, Table~\ref{table:picket}). Therefore, the neutral gas covering fractions are accurately recovered from our observational conditions.  

{The \it James Space Webb Telescope }(JWST) will accurately measure $C_f$ of metal absorption lines from high-redshift galaxies. The JWST's Near-Infrared Spectrograph (NIRSpec) is expected to have $R\sim3000$ (1000) in the high- (medium-) resolution configurations. This means that the systematic errors will be 3\% (6\%) of the measured $C_f$ for S/N~=~5 observations, illustrating the feasibility of measuring $C_f$ from high-redshift galaxies with JWST.

\section{Covering fraction of LyC emitters and comparison sources }
\label{sect:results}

We now examine the derived covering fractions from the Lyman series and the \ion{Si}{ii} absorption lines. Then we discuss different geometrical model assumptions and compare our results to earlier work.

\subsection{Leakers have low neutral gas covering fractions}
\label{result:cfleak}

The \ion{H}{i} covering fraction describes the porosity of the neutral gas and demonstrates whether the neutral gas is clumpy. A smaller $C_f$(\ion{H}{i}) means that there are more low-density channels for ionizing photons to escape through.  

\begin{figure} [t]  
\includegraphics[width=\hsize]{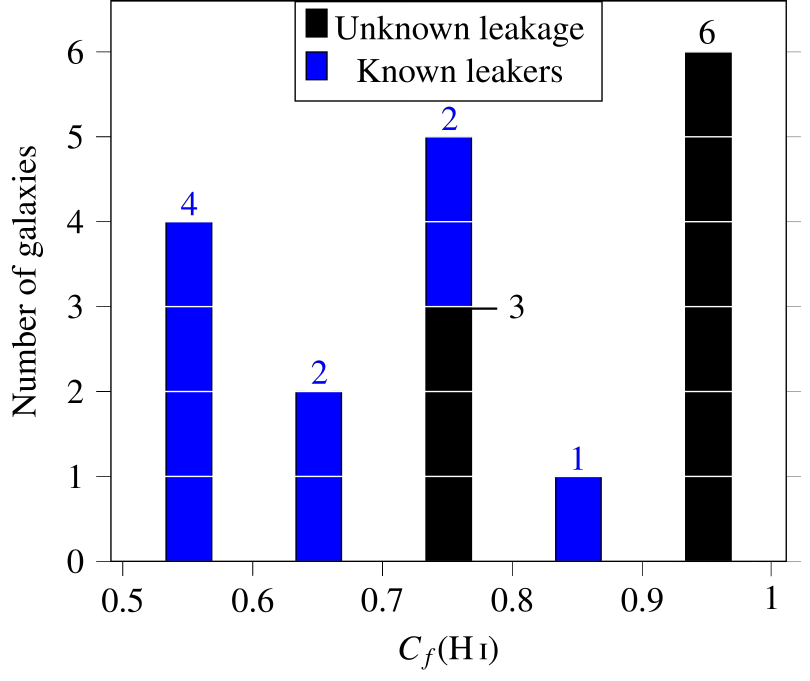}
 \caption{Histogram of the \ion{H}{i} covering fraction ($C_f$(\ion{H}{i})) from the 18 galaxies in our sample. We split the sample into galaxies known to leak LyC photons (blue) and galaxies without measured LyC emission (black). The leakers have the lowest $C_f$(\ion{H}{i}) and the unknown emitters have the highest $C_f$(\ion{H}{i}).}
         \label{fig:histcf}
\end{figure}

Our sample has \ion{H}{i} covering fractions ranging from $0.50$ to unity (Fig.~\ref{fig:histcf}). Only 5 of the 18 galaxies (28\%) have an \ion{H}{i} covering fraction consistent with unity at $1 \sigma$. The low $C_f$(\ion{H}{i}) values are likely because the sample is biased: 15 of these 18 galaxies were targeted as potential LyC leaker or for being particularly strong line emitters. Since a nonunity $C_f$ is a possible  LyC escape mechanism, it is not too surprising that many of these galaxies have low $C_f$(\ion{H}{i}). The three galaxies that were not targeted as LyC leakers (the \megasaura\ galaxies) have $C_f$(\ion{H}{i}) consistent with unity. In contrast, the galaxies with the highest confirmed escape fractions of ionizing photons have the lowest  $C_f$(\ion{H}{i}) values (Fig.~\ref{fig:histcf}). The leakers have a median $C_f$(\ion{H}{i})~$= 0.62 \pm 0.10$, while the unknown leakers have a median $C_f$(\ion{H}{i})~$= 0.95 \pm 0.10$. Lyman continuum emitters have low \ion{H}{i} covering fractions, which allows LyC photons to escape through low-density channels.

The high \nh\ values, estimated from the \ion{O}{i} column densities (Table~\ref{table:oiresult}), further emphasize that ionizing photons must escape through holes in the \ion{H}{i}.  \nh, calculated from $N_{\ion{O}{i}}$, is greater than  $10^{18.6}$~cm$^{-2}$ for the entire sample. Even without converting into \nh, the high $N_\text{\ion{O}{i}}$ values require unphysically high metallicities  (12+log(O/H) > 10) for ionizing photons to escape through low-density regions. At these column densities the Lyman series and Lyman limit are saturated. Ionizing photons cannot pass through the neutral gas unabsorbed. In other words, even LyC galaxies have optically thick \ion{H}{i}; ionizing photons must escape through low-density channels. 

Low $C_f$(\ion{H}{i}) values indicate that the escape of LyC photons is dominated by a patchy ISM or the picket-fence model. However, we find that a low covering fraction is not the only parameter leading to a high \fesc. For example, Mrk~54 has a $C_f$(\ion{H}{i})~$\sim$0.5  and a \fesc\, < 1\%. Dust crucially impacts the LyC escape fraction by removing ionizing photons (see Eq.~\eqref{eq:f_uniform}, which assumes a uniform dust screen). Consequently, the escape fractions cannot simply be inferred from the measured covering fractions, but requires a joint determination of $C_f$ and \ebv\ for a given set of geometrical assumptions (for example cases {\em (a)} or {\em (b)} from in Sect.~\ref{sect:geom}). This is discussed in depth in Sect.~\ref{dis:geometry} and in Paper~II, in which we show that the LyC escape fractions derived using the Lyman series absorption features are consistent with the directly observed escape fractions.

Using Eq.~\eqref{eq:f_uniform}, we can predict which of the nine galaxies in the sample without measured LyC emission should emit ionizing photons. To emit LyC photons, both a low \ebv\ and a low $C_f$ are required. J0926+4427, J1429+0643, and GP1054+5238 are the best candidates in our sample to leak ionizing photons (in order of most likely to emit ionizing photons). Follow-up observations should expect to find \fesc\ values between $0.02-0.06$. 

\subsection{\ion{H}{i} and \ion{Si}{ii} covering fractions}
\label{sect:cf}

\begin{figure} [t]  
\includegraphics[width=\hsize]{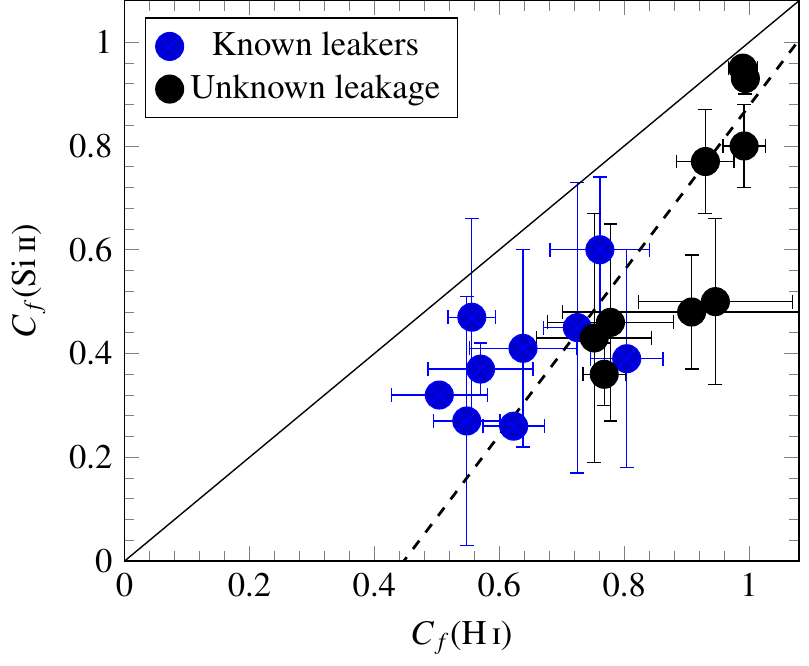}
 \caption{Comparison of the \ion{Si}{ii} and \ion{H}{i} covering fractions. The black solid line shows a one-to-one relation and the dotted line is the fitted relation (Eq.~\eqref{eq:cfhsi}). The average offset between $C_f$(\ion{H}{i}) and $C_f$(\ion{Si}{ii}) is 0.25. 
 }
         \label{fig:hvslis}
\end{figure}

Low-ionization metal absorption lines are often used as proxies for the neutral gas covering fraction \citep{shapley2003, jones2013, alexandroff2015, trainor2015}. Metal lines are in redder portions of the spectra. Consequently, instruments can more efficiently observe metals lines at low redshifts and the Lyman forest does not obscure metal lines at high redshifts. Therefore, metal absorption lines are ideal probes of neutral gas properties  when the Lyman series is not observed. However, recent observations of $z \sim 3$ stacked spectra have suggested that metal absorption lines have covering fractions a factor of two smaller than \ion{H}{i} absorption lines \citep{reddy2016stack}. Here, we test whether $C_f$(\ion{Si}{ii}) traces $C_f$(\ion{H}{i}).

The $C_f$(\ion{Si}{ii}) is systematically lower than $C_f$(\ion{H}{i}) and has a mean offset of $0.25$ (Fig.~\ref{fig:hvslis}). $C_f$(\ion{H}{i}) and $C_f$(\ion{Si}{ii}) are linearly related at the $3\sigma$ significance level (p-value < 0.001; Pearson's correlation coefficient of 0.79) as
\begin{equation}
    C_f(\ion{H}{i}) = (0.63 \pm 0.12) \times C_f(\ion{Si}{ii}) + (0.44 \pm 0.07) .
    \label{eq:cfhsi}
\end{equation}
While $C_f$(\ion{Si}{ii}) is not equal to $C_f$(\ion{H}{i}), Eq.~\eqref{eq:cfhsi} estimates $C_f$(\ion{H}{i}) from $C_f$(\ion{Si}{ii}). This empirically derived relation estimates $C_f$(\ion{H}{i}) from measurements of the metal line covering fraction (here from \ion{Si}{ii} lines), even if the Lyman series lines are not accessible. This emphasizes a clear practical advantage of the uniform dust screen geometry: the stellar attenuation can be estimated without directly measuring $C_f$(\ion{H}{i}). The stellar attenuation then defines the continuum level from which $C_f$(metal) can be estimated and, in turn, C$_f(\ion{H}{i})$ estimated from Eq.~\ref{eq:cfhsi}. Consequently, C$_f(\ion{H}{i})$ can be estimated using C$_f(\ion{Si}{ii})$ and stellar continuum fits of high-redshift galaxies, where the Lyman series is not directly observable. This is not possible in
 the clumpy geometry because it requires $C_f$(\ion{H}{i}) to fit the stellar continuum. In Paper II we used the Lyman series absorption lines and dust attenuation to reproduce the observed Lyman continuum escape fraction.

There are numerous reasons why $C_f$(\ion{Si}{ii}) could be related to, but not equal to, $C_f$(\ion{H}{i}) \citep[see][for an in-depth discussion]{reddy2016stack}. First, narrow, high covering fraction, metal lines may be unresolved in low-resolution spectra ($\sim$1500). While this is possible for low-resolution data, it is less likely for the high-resolution COS data. The \ion{Si}{ii} lines could possibly be unsaturated, but the doublet method accounts for this possibility and does not remove the systematic offset. Alternatively, the \ion{Si}{ii} ionization potential overlaps with, but is not equal to, \ion{H}{i}. Therefore, the \ion{Si}{ii} and \ion{H}{i} gas may trace similar, but not equal, gas. However, \citet{reddy2016stack} have found that the \ion{Si}{ii} and \ion{H}{i} line profiles and central velocities are similar \citep[also see][]{chisholm2016}. This indicates that the two transitions are comoving and trace similar gas. Finally, the \ion{Si}{ii} lines are also potentially affected by scattering and fluorescent emission in-filling, although it is difficult to predict whether the amount the emission increases the covering fraction \citep[cf.][]{prochaska2011, scarlata2015}.

 Another explanation is that the neutral gas and metal-enriched gas are not fully mixed. In this situation, \ion{Si}{ii} only probes metal-rich regions, while the \ion{H}{i} gas probes both high and low metallicity gas. Therefore, the neutral gas covers more of the background source than the metals alone. At lower metallicities this is exaggerated because there is less metal-rich gas. This leads to fewer high-density metal regions to absorb the background continuum, and systematically reduces the metal covering fraction. To test the effect of metallicity on the relation between the neutral and metal covering fractions, we fit a multiple linear relationship between $C_f$(\ion{H}{i}), $C_f$(\ion{Si}{ii}), and \oh. We find a significant  (3$\sigma$, p-value < 0.001, Pearson's correlation coefficient of 0.79) trend between these three parameters that scales as
\begin{multline}
    C_f(\ion{H}{i}) = \left(1.8\pm0.8\right) - \left(0.18\pm0.10\right)\times \left[12+\log(\text{O/H})\right] \\
    +\left(0.75 \pm 0.16\right) \times C_f(\text{\ion{Si}{ii}}).
    \label{eq:cfhimet}
\end{multline}
While Eq.~\ref{eq:cfhimet} is only marginally more significant than the simpler relation between $C_f$(\ion{Si}{ii}) and $C_f$(\ion{H}{i}), it physically explains the relation between the neutral and metal covering fractions.

The difference between Eq.~\eqref{eq:cfhsi} and Eq.~\eqref{eq:cfhimet} is the metallicity dependence of the correction added to $C_f$(\ion{Si}{ii}). As the metallicity increases, the required correction to convert $C_f$(\ion{Si}{ii})  into $C_f$(\ion{H}{i}) decreases. While at lower metallicities, the factor added to $C_f$(\ion{Si}{ii}) must be larger. Physically, this means that at lower metallicities the \ion{Si}{ii} traces a smaller fraction of the total \ion{H}{i}. Therefore, the linear relationship between $C_f$(\ion{H}{i}) and $C_f$(\ion{Si}{ii}) may arise because metal-rich clumps cover the background source differently than the \ion{H}{i} gas.

The empirical relationship, Eq.~\eqref{eq:cfhsi}, is recovered from Eq.~\eqref{eq:cfhimet} when the median metallicity of the sample (8.06) is used. However, Eq.~\eqref{eq:cfhsi} would change if the median \oh\ of the sample changes. This suggests that the empirical relationship between $C_f$(\ion{H}{i}) and $C_f$(\ion{Si}{ii}) is only constrained near the metallicities of the sample. For divergent \oh\ values, the metallicity should be accounted for when estimating $C_f$(\ion{H}{i}).

\subsection{Is the dust geometry clumpy or uniform?} 
\label{dis:geometry}
\begin{figure}[t]
\includegraphics[width=\linewidth]{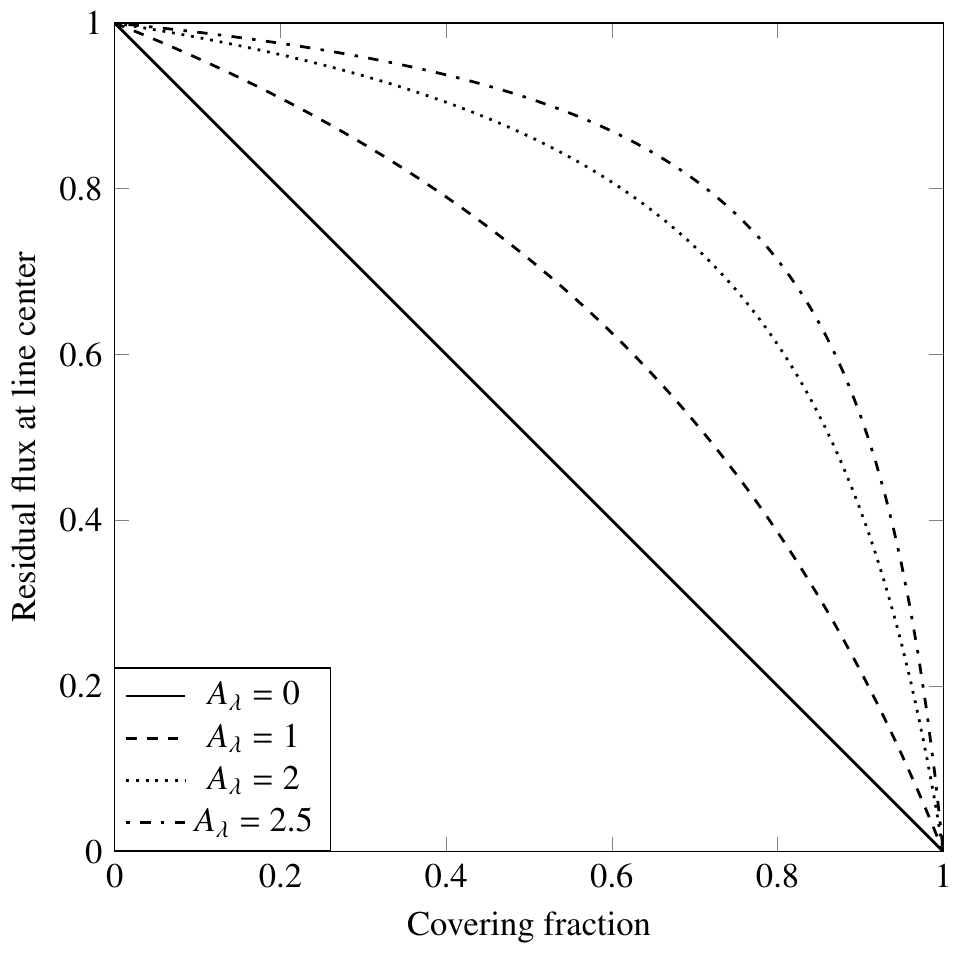}
\caption{Residual flux $R$ of a saturated absorption line from Eq.~\eqref{eq:d_holes} for various values of $A_\lambda = k_\lambda \ebv$ in a clumpy ISM geometry. When dust is present, the residual flux is higher
than $(1-C_f)$. However, in a uniform dust screen geometry, the residual flux is always $R=(1-C_f)$. The covering fraction derived from the observed residual flux and a clumpy dust geometry is underestimated when dust is present.}
\label{fig:cfdepth}
\end{figure}
     
In Sect.~\ref{result:cfleak}, we emphasized that dust heavily contributes to the escape of ionizing photons. However, since the impact of dust depends on the assumed geometry, we explore the two cases described above (Sect.~\ref{sect:geom}), namely {\em (a)} a uniform dust screen and {\em (b)} a clumpy geometry dust model. To test the impact of these assumed geometries on $C_f$ and \ebv, we modify our fitting routine from {\em (a)} a uniform dust screen model to  {\em (b)} a clumpy model. We refit two galaxies: J0921+4509 and J1152+3400, which have an average $C_f$(\ion{H}{i}) and low $C_f$(\ion{H}{i}), respectively. Adopting a clumpy geometry {\em (b)}, we derive \ebv $=0.236$ and 0.239 (versus \ebv\ = 0.224 and 0.134 in model {\em (a)}) and $C_f$(\ion{H}{i})~$=0.976$ and 0.912 (versus $C_f$ = 0.754 and 0.506  in model {\em (a)}), respectively. These values and Eq.~\eqref{eq:f_holes} predict a LyC escape fraction of $\fesc=0.024$ and 0.088 at $\lambda = 912~\AA$, which is consistent, within $1\sigma$, with the values derived using the uniform dust screen model ($\fesc=0.017 \pm 0.007$\% and $0.092 \pm 0.027$\%). 

As just shown, using the uniform dust screen leads to a lower \ebv\ and $C_f$(\ion{H}{i}) compared to the clumpy model. In the clumpy model, all of the light that escapes through holes in the neutral gas is unattenuated. Therefore, the holes in the neutral gas must be smaller, while any continuum passing through the neutral gas must be more heavily attenuated to match the observed flux. $C_f$(\ion{H}{i}) and \ebv\ must take values that fit both the geometry and the observed flux (extending to Lyman continuum). Both quantities are therefore model dependent. However, disentangling the actual geometric model is not mandatory to predict \fesc\ with $C_f$(\ion{H}{i}) and \ebv, as long as the geometry is consistently modeled (see Paper II).

On the other hand, the observed residual flux of the \ion{H}{i} absorption lines provide clues to the geometric model. Eq~\eqref{eq:d_holes} shows that the residual flux of the \ion{H}{i} absorption lines are wavelength dependent in the clumpy geometry. Therefore, if the residual flux of the individual Lyman series absorption lines changes with wavelength (for example from Ly$\beta$ to Ly6), then the data would suggest a clumpy dust geometry. Fig.~\ref{fig:cfdepth} shows the relationship between the residual flux of a saturated absorption line (R) and the neutral covering fraction for various values of $A_\lambda = k_\lambda \ebv$ in a clumpy geometry. As $A_\lambda$ increases (larger \ebv\ or a bluer transition), the residual flux decreases as dust removes more continuum photons. In clumpy dust geometry, R is less than 1-$C_f$ if there is dust. For a covering fraction of 0.8, the measured R is 0.2 when $A_\lambda$ is 0, 0.4 when $A_\lambda$ is 0.1, and 0.7 when $A_\lambda$ is 0.2 (Fig.~\ref{fig:cfdepth}). The $C_f$ estimated from 1-R is underestimated in the clumpy dust geometry case.

\begin{table}
    \caption{Theoretical residual flux for the Lyman series (Ly$\beta$ to Ly6) of a galaxy with a clumpy dust geometry.}             
    \label{table:depthbias}  
    \centering          
    \begin{tabular}{c c c c c c c c }     
    \hline\hline       
    \multirow{2}{*}{\ion{H}{i} lines}&\multicolumn{4}{c}{\ebv }  \\ 
     & 0.0& 0.1 & 0.2 & 0.3  \\ \hline
      Ly$\beta$ & {0.100}& 0.245 & 0.489 & 0.737  \\ 
      Ly$\gamma$ &{0.100}  &0.254 & 0.512 & 0.764  \\ 
      Ly$\delta$ &{0.100} &0.259 &  0.523 &  0.775  \\ 
      Ly5 &{0.100} &0.261 & 0.530 & 0.782  \\
      Ly6   &{0.100} &0.262 &  0.533 &  0.785   \\
    \hline                  
    \end{tabular}
    \tablefoot{Theoretical measurements of the residual flux of a saturated absorption line for five Lyman series lines. These values are calculated assuming \fesc~$=0.10$, a clumpy dust geometry, saturated Lyman lines, and the \citet{reddy2016dustlaw} dust attenuation law. The differences found between various Lyman lines for a given \ebv\ are small and accurate residual flux measurements are required to distinguish the geometries.}
    \end{table}

Since the residual intensity increases with $A_\lambda$ and $k_\lambda$, this effect, if present, should be detectable in the Lyman series absorption lines. Nevertheless, the differences between the residual flux of the Lyman series should be small and most easily detected in galaxies with large \ebv. Table~\ref{table:depthbias} shows the expected residual flux of saturated Lyman series lines, assuming an \fesc~$=0.10$ ($C_f$ = 0.9 in the  clumpy geometry). When \ebv\, is 0.1, the difference between the Ly$\beta$ and Ly6 residual flux is less than 0.02. For \ebv = 0.30, the offset rises to 0.05. While we do not measure a trend in our observations (Table~\ref{table:cfdepth}), the measured residual intensities have large uncertainties. Our simulations imply that S/N~$=30$ and R~$=15000$ are required to distinguish the two geometries. We cannot determine the dust geometry with the current data and both a uniform dust screen or clumpy geometry are allowed by the observations.

\subsection{Comparison with other studies}
\label{sect:comp}

\citet{reddy2016stack} recently determined the \ion{H}{i} covering fraction of $z \sim 3$ Lyman break galaxies from Lyman series fits. Analyzing stacked spectra, they found high $C_f(\ion{H}{i}) \sim 0.92-0.97$. They also examined the relation between $C_f$(\ion{H}{i}) and the $C_f$ of metal lines (\ion{Si}{ii}, \ion{C}{ii}, and \ion{Al}{iii}), finding a much lower covering fraction for the metals than for \ion{H}{i}. The differences between their $C_f$(metal) and their $C_f$(\ion{H}{i}) are nearly twice as large as we find for $C_f$(\ion{Si}{ii}) in Fig.~\ref{fig:hvslis}  (their $C_f(\rm metal)$ is 0.4-0.6 smaller than $C_f$(\ion{H}{i}), versus our average of 0.25).
 
This discrepancy arises from an inconsistency in determining $C_f$(metal). To fit their $C_f$(\ion{H}{i}), \citet{reddy2016stack} adopted a clumpy geometry with dust only in the \ion{H}{i}. The continuum level used to obtain the normalized composite spectra in their study is determined as
\begin{equation}
    F_\lambda^\text{cont} = F_\lambda^\star 10^{-0.4 \text{\ebv} k_\lambda} \times C_f(\text{\ion{H}{i}}) + F_\lambda^\star [1-C_f(\text{\ion{H}{i}})]
.\end{equation}
In this geometry, the residual flux of a saturated metal absorption line is
\begin{equation}
    R(\text{metal}) = \frac{F_\lambda}{F_\lambda^\text{cont}} = \frac{1-C_f(\text{metal})}{10^{-0.4\text{\ebv}k_\lambda} \times C_f(\text{\ion{H}{i}}) +1-C_f(\text{\ion{H}{i}})}
    \label{eq:reddy}
.\end{equation}
We note that Eq.~\eqref{eq:reddy} reduces to Eq.~\eqref{eq:d_holes} when $C_f$(metal)~$=C_f$(\ion{H}{i}).  We stress that the continuum normalization impacts the determination of $C_f$ from R because R depends on the assumed dust geometry. In the uniform dust screen geometry (a) (the geometry that we predominately use, i.e., in Fig~\ref{fig:histcf}), the covering fraction is simply 1-R. Conversely, in the clumpy model (b), the  covering fraction of a saturated absorption lines is {\em not} simply 1-R. The covering fraction cannot be read off from the profile; a measured \ebv\ is required to determine $C_f$ \citep[see the large differences in Fig.~\ref{fig:cfdepth};][]{vasei2016}. However, \citet{reddy2016stack} measured $C_f(\rm metal)$ as $1-R$ from the continuum normalized spectra. Meanwhile, they determined $C_f(\ion{H}{i})$ from the full fits of their spectra, which properly accounts for their assumed geometry. This means that $C_f(\rm metal)$ and $C_f(\ion{H}{i})$ are calculated with different geometric assumptions.

\begin{figure}[t]
\includegraphics[width=\linewidth]{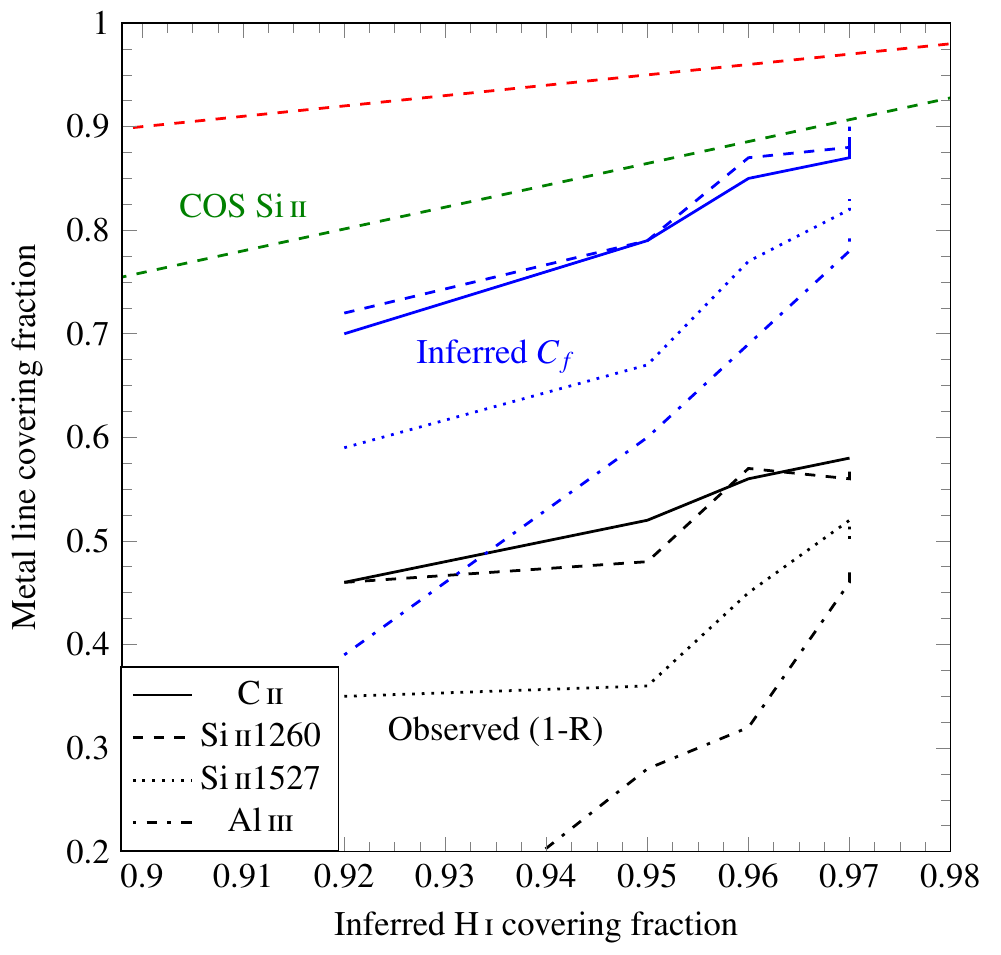}
\caption{Metal covering fractions from \ion{Si}{ii}, \ion{C}{ii,} and \ion{Al}{iii} (as given by the line types in the legend) as a function of the \ion{H}{i} covering fraction ($C_f(\ion{H}{i})$) from \citet{reddy2016stack}. $C_f$(\ion{H}{i}) is always measured in a clumpy geometry. The black lines show $1-R$ measured from the dereddened stellar continuum normalized stacks. This assumes a clumpy geometry with a uniform dust screen. The blue lines show when the $C_f(\rm metal)$ is derived in the same geometry as $C_f(\ion{H}{i})$. When calculated with consistent geometries, $C_f$(\ion{Si}{ii}~1260\AA) (blue dashed line) is offset from $C_f$(\ion{H}{i}) by a similar amount as our observations in Fig.~\ref{fig:hvslis}. The green dashed line shows our fitted relationship between $C_f(\ion{Si}{ii})$ and $C_f(\ion{H}{i})$ from Fig.~\ref{fig:hvslis}, after converting it to clumpy geometry. The \citet{reddy2016stack} $C_f$(\ion{H}{i})-$C_f$(\ion{Si}{ii}) relationship is consistent, within the errors, with ours. The red dashed line shows a one-to-one relation.}
\label{fig:reddy}
\end{figure}

The inconsistency in the derivation of $C_f$(metal) and $C_f$(\ion{H}{i}) produces the large mismatch between our observations and \citet{reddy2016stack}. We correct the observed residual flux from \citet{reddy2016stack} to recompute consistently
the $C_f(\rm metal)$ values using Eq.~\eqref{eq:reddy} and their measured \ebv\ (using their SMC extinction law) and $C_f$(\ion{H}{i}) values (Fig.\ \ref{fig:reddy}). The large offset between the \ion{Si}{ii}~1260\AA\ and \ion{H}{i} is halved (compare the dashed blue line to the dashed black line), although $C_f(\rm metal)$ is still systematically smaller by $\sim 0.2$. This offset is similar to the offset that we measure in Fig.~\ref{fig:hvslis} (green dashed line in Fig.~\ref{fig:reddy}). The discrepancy between $C_f$(metal) and $C_f$(\ion{H}{i}) emphasizes that $C_f$ can only be read off from the residual flux of absorption lines in the uniform dust screen geometry because this continuum normalization accounts for the dust attenuation. In other geometries, the dust attenuation must be accounted for when measuring $C_f$.

The small remaining offset ($\sim 0.2$) between the $C_f$(\ion{H}{i}) and $C_f$(metal) is comparable to the $C_f$(\ion{Si}{ii})-$C_f$(\ion{H}{i}) relation found in Eq.~\eqref{eq:cfhsi}. To demonstrate this, we convert our $C_f$(\ion{H}{i}) and $C_f$(\ion{Si}{ii}) values, measured in a uniform geometry, to a clumpy geometry, and recalculate how $C_f$(\ion{Si}{ii}) scales with $C_f$(\ion{H}{i}) (green line in Fig.~\ref{fig:reddy}). This recalculated relation is within the errors of the \citet{reddy2016stack} \ion{Si}{ii}~1260\AA\ relation (blue dashed line). The consistent scaling between $C_f$(\ion{H}{i}) and $C_f$(\ion{Si}{ii}) from our study and \citet{reddy2016stack} demonstrates first that  $C_f(\text{\ion{Si}{ii}}) \neq C_f(\text{\ion{H}{i}})$, and second that empirical relations can convert $C_f$(\ion{Si}{ii}) into $C_f(\ion{H}{i})$ for a variety of galaxy types. Both findings are crucial for indirect measurements of the Lyman continuum escape fraction, as discussed in Paper II.

There is still a $C_f$ spread among the different transitions: strong \ion{Si}{ii}~1260\AA\ and \ion{C}{ii} lines have similar $C_f$, but weaker \ion{Si}{ii}~1527 and \ion{Al}{iii} lines have smaller $C_f$. Puzzlingly, the \ion{Si}{ii}~1260\AA\ and \ion{Si}{ii}~1527\AA\ lines arise from the same transition, and should have the same $C_f$. Different $C_f$(\ion{Si}{ii}~1260\AA) and $C_f$(\ion{Si}{ii}~1527\AA) would be expected if \ion{Si}{ii}~1527\AA\ was not saturated, but \citet{reddy2016stack} showed that the \ion{Si}{ii}~1527\AA/\ion{Si}{ii}~1260\AA\ equivalent width ratio is unity. Consequently, both transitions are saturated. Alternatively, since  \ion{Si}{ii}~1260\AA\ is a stronger line than \ion{Si}{ii}~1527\AA\ ($f$ value of 1.22 versus 0.133), there could be more optically thick \ion{Si}{ii}~1260\AA\ regions along the line of sight than optically thick \ion{Si}{ii}~1527\AA\ regions. Intersecting more regions along the line of sight covers more of the background continuum, such that $C_f$(\ion{Si}{ii}~1260\AA) is larger than $C_f$(\ion{Si}{ii}~1527\AA). This differential covering fraction is similar to the proposed physical mechanism creating the difference between  $C_f(\text{\ion{Si}{ii}})$ and  $C_f(\text{\ion{H}{i}})$ in Sect.~\ref{dis:geometry}. In the \ion{Si}{ii}~1260\AA\ and \ion{Si}{ii}~1527\AA\ case, the number of absorbers is different owing to differences in line strength, not metallicity.

\section{Conclusions}
\label{sect:conc}

We have analyzed the \ion{H}{i}, \ion{O}{i}, and \ion{Si}{ii} low ionization interstellar  absorption lines from a sample of 18 star-forming galaxies with rest-frame ultraviolet spectroscopy of the Lyman series. The majority of the sources have COS spectra taken with the HST and are at low redshift ($z < 0.4$). Our sample includes nine Lyman continuum leaking galaxies. We fit the stellar continuum, dust attenuation, \ion{H}{i} Lyman series absorption lines, and several low ionization absorption lines (\ion{Si}{ii} $\lambda\lambda$1190, 1193 and $\lambda$1260 \AA, \ion{O}{i} $\lambda$ 1039 \AA\ in particular). These fits determine the UV attenuation, as well as column densities and covering fractions of neutral hydrogen and metals (Sect.~\ref{sect:method}). Additionally, we applied our fitting method to synthetic Lyman series ISM absorption lines to investigate the systematic errors of the covering fractions and column densities. Then, we studied the observed Lyman series lines to constrain the \ion{H}{i} properties. The direct \ion{H}{i} properties were compared to indirect estimates of the neutral gas properties using \ion{O}{i} and \ion{Si}{ii} absorption lines. 

Our main results are summarized as follows:
\begin{itemize}
\item The \ion{H}{i} covering fraction is accurately recovered from the Lyman series. Synthetic spectra recover the covering fractions with low systematics for a wide range of S/N and resolutions (S/N $\geq 2$ and  $R > 3000$, or S/N $\geq 5$ and $R \geq$ 600; Fig.~\ref{figure:picket}). Future observatories, such as JWST, will accurately measure covering fractions of high-redshift galaxies.

\item The observed \ion{H}{i} lines are found to be saturated in all galaxies. Assuming the same O/H abundance in the neutral and ionized gas, we derive \ion{H}{i} column densities  of $\log(\nh) \sim 18.6-20$  cm$^{-2}$ from the \ion{O}{i} absorption lines (Sect.~\ref{sect:oi}).

\item  The \ion{H}{i} column densities derived for the known LyC leakers are too high to allow ionizing photons to escape. Rather, we find that the LyC emitting galaxies have \ion{H}{i} covering fractions below unity. Ionizing photons escape through optically thin holes in a clumpy ISM. The median \ion{H}{i} covering fraction of confirmed LyC emitting galaxies is 0.62, as compared to 0.95 for galaxies that do not have LyC detections.

\item The \ion{Si}{ii} covering fraction is systematically lower than the \ion{H}{i} covering fraction (Fig.~\ref{fig:hvslis}). However, the \ion{Si}{ii} covering fraction is found to scale linearly with the \ion{H}{i} covering fraction (Eq.~\eqref{eq:cfhsi}). We show that this relation is compatible with the relationship from  \cite{reddy2016stack} of stacked $z \sim 3$ spectra. Thus, with an empirical correction, the \ion{Si}{ii} absorption lines can be used to determine the  \ion{H}{i} covering fraction when the Lyman series is not observable. This is especially powerful at high redshift when the Lyman series cannot be observed.

\item The assumed dust geometry (here a uniform screen) impacts the measured covering fractions and dust attenuations (Fig.~\ref{fig:cfdepth}), but it does not impact the inferred escape fractions of ionizing photons (Sect.~\ref{dis:geometry}). Crucially, the geometric covering fraction cannot be read off from the residual line flux in all geometries. A consistent fitting of the dust attenuation, continuum, and absorption lines is required to determine the covering fraction properly.
 \end{itemize}
 
By relating the \ion{H}{i} covering fraction to metal absorption lines, we have provided the framework to measure the neutral gas covering fractions in star-forming galaxies. This will be of particular interest at high redshift, where the Lyman series and Lyman continuum are not observable and indirect measurements are required to measure the escape fraction of ionizing photons. In a companion paper \citep{chisholm2018}, we use the absorption lines to accurately predict the escape fraction of the known LyC leakers. These methods also yield consistent results for a sample of $z \sim 2-2.4$ galaxies. Our analysis emphasizes that UV spectra of sufficient quality (S/N $>$ 5) from JWST and extremely large telescopes may constrain how the Universe became reionized.

\begin{acknowledgements}
We would like to thank the anonymous referee for carefully reading our manuscript and for giving constructive comments which improved the quality of the paper.
\end{acknowledgements}

\bibliographystyle{aa}

\begin{appendix}

\section{Simulated spectra}
\label{sect:data}
  \label{step1}

Here we detail the procedure we used to create the simulated spectra in Sect.~\ref{meth:simu}. We simulated synthetic spectra to mimic the observed Lyman series. Therefore, we created synthetic spectra using a single {\small STARBURST99} stellar continuum model of R(S99) $\approx$ 2500 \citep{leitherer1999, leitherer2010}  with a stellar age of 3 Myr, a solar metallicity, and without dust attenuation. This stellar age has a strong \ion{O}{iv} P-Cygni profile (broad blueshifted absorption and redshifted emission) that blends with the Ly$\beta$ absorption line (see Fig.~\ref{fig:lybzoom}). The synthetic spectra include the wavelength regime between 910-1050~$\angstrom$ with a pixel separation of 0.02~$\angstrom$ (6 km/s at 1035~$\angstrom$). These wavelengths include  the Lyman series, from Ly$\beta$ to the Lyman break ($\approx$ 911.8~$\angstrom$). We imprinted on top of the stellar continuum  ISM absorption lines from \ion{H}{i}, \ion{O}{i}, \ion{O}{vi,} and  \ion{C}{ii} (see \autoref{table:lines}) with Voigt profiles. The Doppler broadening parameters ($b$), column densities ($N$), and covering fractions ($C_{f}$) are created with independent values for each element (see Table~\ref{table:species}). We fixed the velocities of each line at 0~km~s$^{-1}$. The Einstein-A coefficients and the oscillator strengths of each transition were taken from the NIST Atomic Database\footnote{\url{https://physics.nist.gov/PhysRefData/ASD/lines_form.html}}.

    \begin{table}
    \caption{Absorption lines, wavelengths, and parameters used to create the synthetic absorption lines. } 
    \label{table:species} 
    \centering                          
    \begin{tabular}{ccccc}
    \hline \hline 
            (1) &  (2) & (3) & (4) & (5) \\  
       Ion & $v$  & $b$ & $\log(N)$ & $C_{f}$\\ 
       & [km s$^{-1}$] & [km s$^{-1}$]&[cm$^{-2}$]  &\\  \hline 
        \multirow{2}{*}{\ion{H}{i}}  &   \multirow{2}{*}{0}  & \multirow{2}{*}{50}  &   17.57\tablefootmark{a}  & 1\tablefootmark{a} \\ 
       &  &  & 20.00\tablefootmark{b} & 0.9\tablefootmark{b} \\
      \ion{O}{i} & 0 &  75 & 16 & 1 \\
     \ion{O}{vi} & 0 &  100 & 15 & 1 \\
        \ion{C}{ii} &0 & 80 & 15.5 &1\\ \hline 
    \end{tabular}
    
    \tablefoot{(1) Ion; (2) velocity shift of the lines; (3) Doppler parameter; (4) logarithm of the column density; (5) covering fraction. \\
    \tablefoottext{a}{Denotes values used for the simulated density-bounded spectra} \\
    \tablefoottext{b}{Denotes values used for the simulated picket-fence spectra} }         
    \end{table}

We defined two categories of synthetic spectra corresponding to an ionizing escape fraction of 10\%. This value is drawn from the largest known publicly available low-redshift LyC escape fractions \citep{izotov2016a, izotov2016b}. For the first category, we assumed holes in an otherwise optically thick ISM allow ionizing photons to escape. We called this the picket-fence model. For the picket-fence model, the \ion{H}{i} parameters are N$_{\ion{H}{i}} = 10^{20}$ cm$^{-2}$ ($\e^{-\tau_{\mathrm{H}}} \rightarrow 0$) and $C_f(\mbox{\ion{H}{i}}) = 0.90$. For the second scenario, we assumed a low \ion{H}{i} column density allows ionizing photons to escape (the density-bounded regime). We fixed N$_{\ion{H}{i}} = 10^{17.57}$ cm$^{-2}$ ($\e^{-\tau_{\mathrm{H}}} \approx 0.10$) and $C_f(\mbox{\ion{H}{i}}) = 1.0$. Since we fixed \ebv\ = 0 in all our simulations, the geometry does not change our simulations (Sect.~\ref{sect:geom}).

To reproduce the observations, we created synthetic spectra at various instrumental resolutions and S/N ratios. We simulated seven resolutions, $R =\lambda/\delta\lambda$ (where $\delta\lambda$ is the FWHM of the spectra), of 15000, 6000, 3000, 1500, 1000, 750, and 600. For the synthetic spectra having R > R(S99), this is performed by combining the theoretical stellar continuum with R(S99) = 2500 with absorption lines of spectral resolution R. The range of resolutions chosen correspond to spectrographs such as  the COS on HST, the Low and High Resolution Imaging Spectrometer (LRIS/HRIS) on Keck, MagE on Magellan, Muse, or the upcoming NIRSpec on  JWST. For each R, we generated seven S/N ratios (2, 5, 10, 20, 30, 40, and 50) in a sufficiently large range to study typically delivered noise levels. At each S/N, we generated 50 different sets of random Gaussian noise and added the noise to the same synthetic spectra. This produces a sample of 50 synthetic spectra where only the random noise varies between the spectra. Figures~\ref{fig:singlefit15000} and~\ref{fig:singlefit600} show simulated spectra with resolutions of 15000 and 600 for three S/N ratios (noise-free, 10 and 2). These results are discussed in Sect.~\ref{sect:sims}.

\section{Fits}

\begin{figure*}
   \centering
   \includegraphics[scale = 0.55]{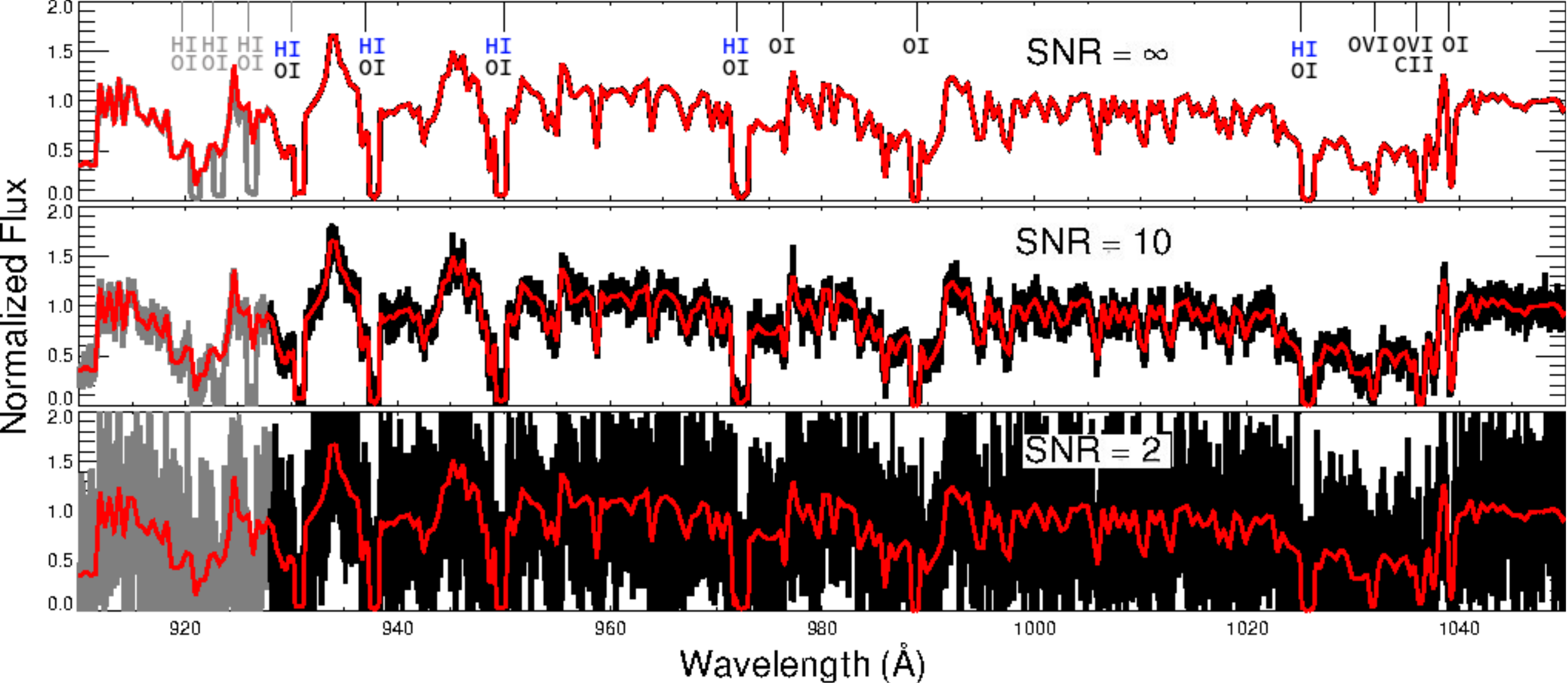}
     \caption{Fits (in red) of the synthetic spectra at high resolution (R = 15000) for the picket-fence model ($C_f$(\ion{H}{i}) $= 0.9$, \nh$= 10 ^{20} $cm$^{-2}$) with S/N equal to $\infty$ (top), 10 (middle) and 2 (bottom). We exclude the gray portion when fitting. In the upper portion of the top panel, we indicate and label the fitted absorption lines.}
         \label{fig:singlefit15000}
   \end{figure*}

\begin{figure*}
   \centering
   \includegraphics[scale = 0.55]{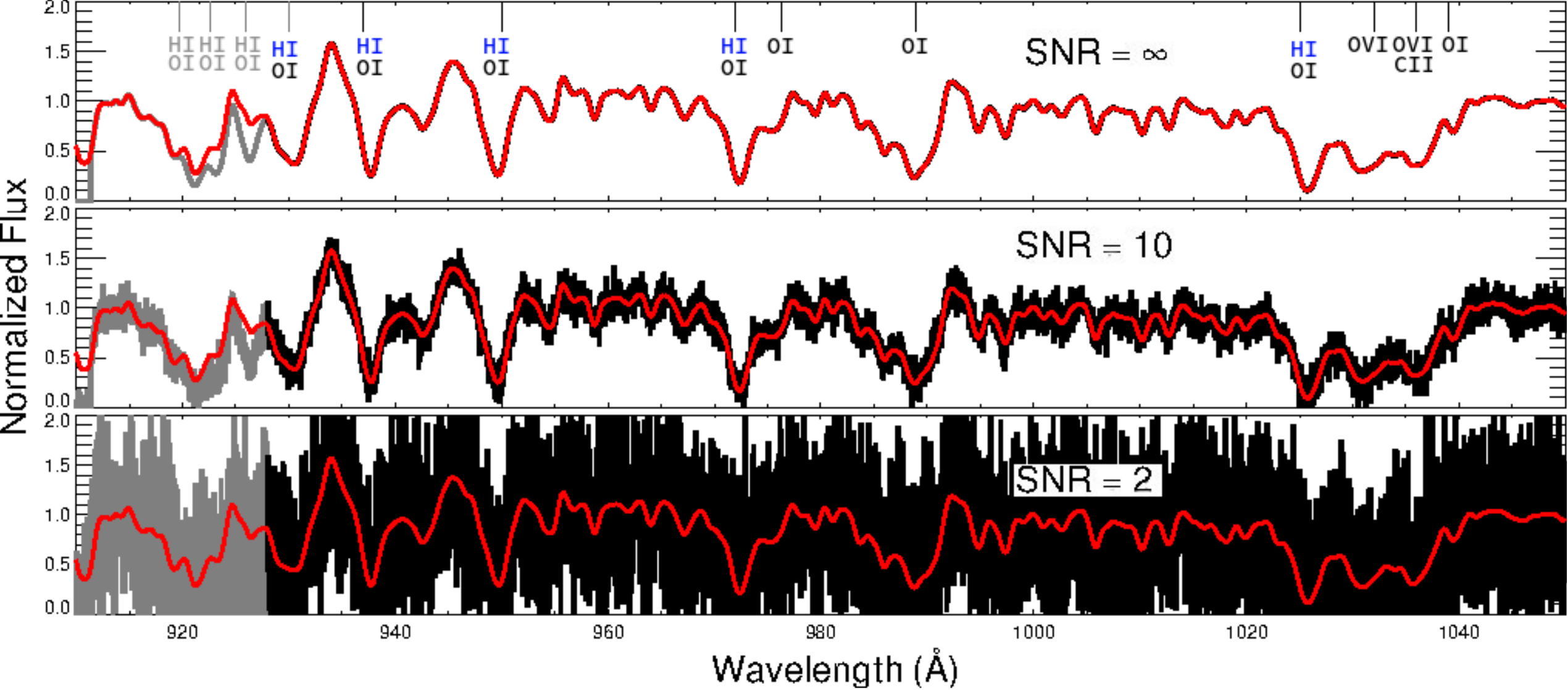}
     \caption{Same as Fig.~\ref{fig:singlefit15000} but for a spectral resolution of R = 600.}
         \label{fig:singlefit600}
   \end{figure*}

\begin{figure*}
   \centering
   \includegraphics[scale = 0.55]{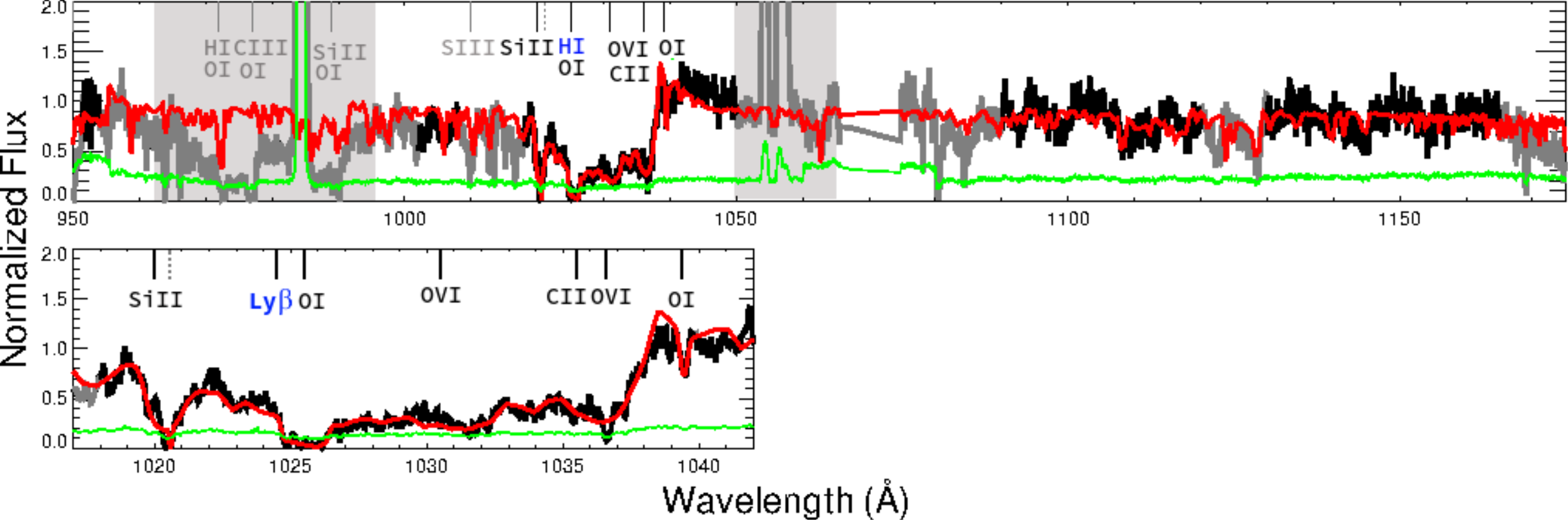}
     \caption{Fit (in red solid line) of the COS G130M spectrum of the galaxy J0921+4509. Black is the observed flux included in the routine either to fit the stellar continuum or the ISM absorption lines. Gray portions are masked out for both steps; gray shaded regions indicate those masked because of geocoronal emission. The flux error array appears in green. We display the ISM and Milky Way absorption lines as solid and dotted lines in the upper portion of each panel. Black or blue labels indicate that the lines are fit, whereas gray labels indicate that they are not fit. When present, red labels indicate lines that are not detected.
     Reference for this observation: \citet{borthakur2014}}
         \label{fig:J0921}
   \end{figure*}

\begin{figure*}
   \centering
   \includegraphics[scale = 0.55]{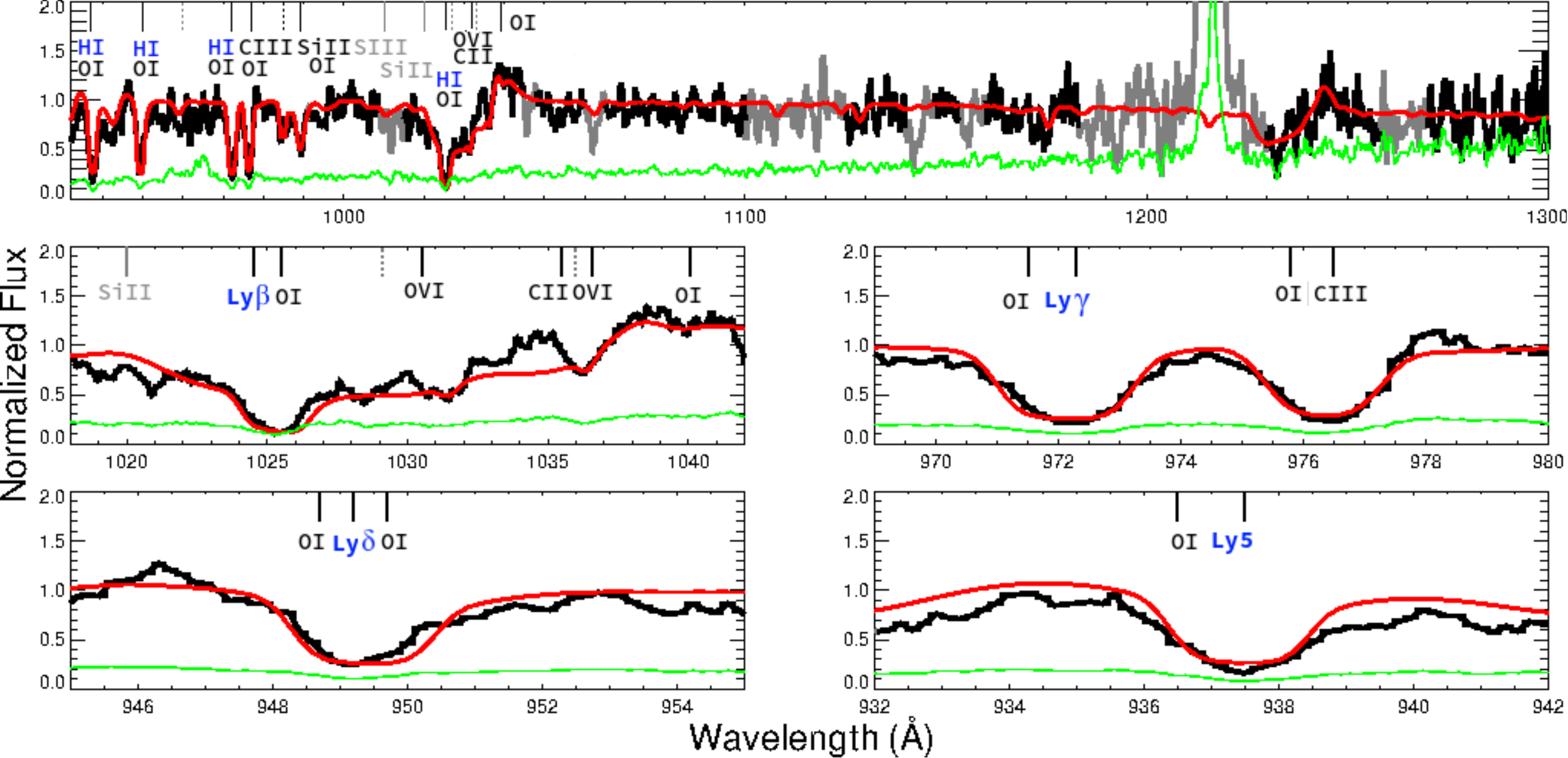}
     \caption{Same as Fig.~\ref{fig:J0921} but for the COS G140L spectrum of J1503+3644. Reference for this observation: \citet{izotov2016a}}
         \label{fig:J1503}
   \end{figure*}

\begin{figure*}
   \centering
   \includegraphics[scale = 0.55]{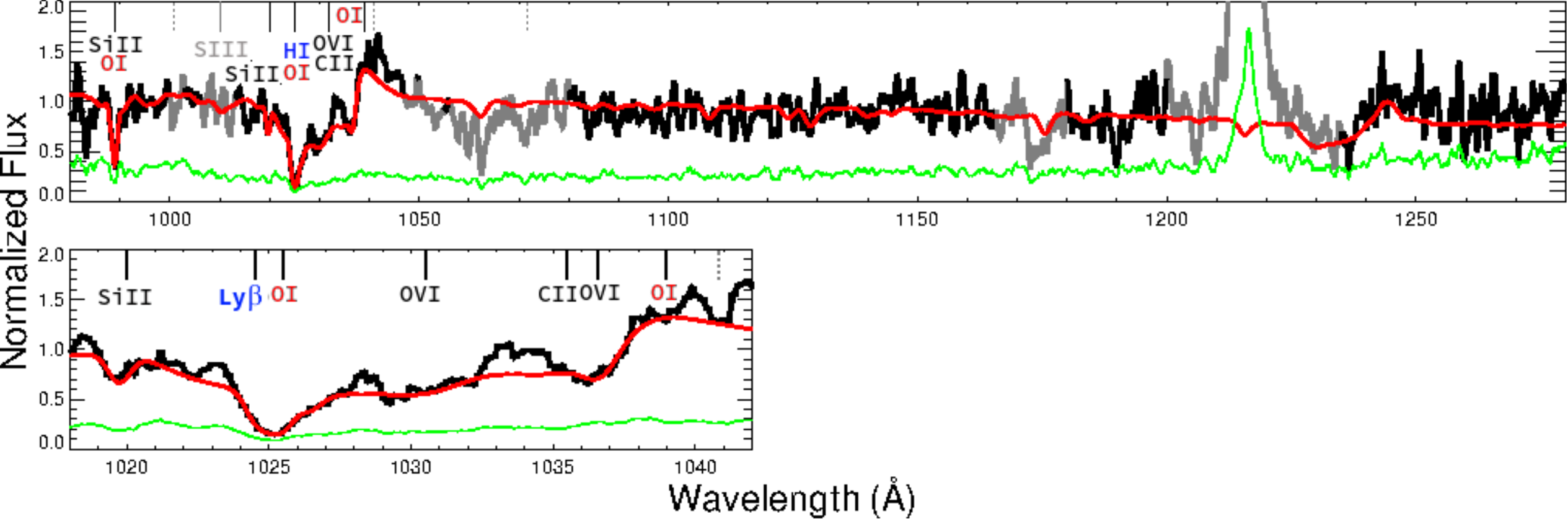}
     \caption{Same as Fig.~\ref{fig:J0921} but for the COS G140L spectrum of J0925+1409. 
     Reference for this observation: \citet{izotov2016b}}
         \label{fig:J0925}
   \end{figure*}

\begin{figure*}
   \centering
   \includegraphics[scale = 0.55]{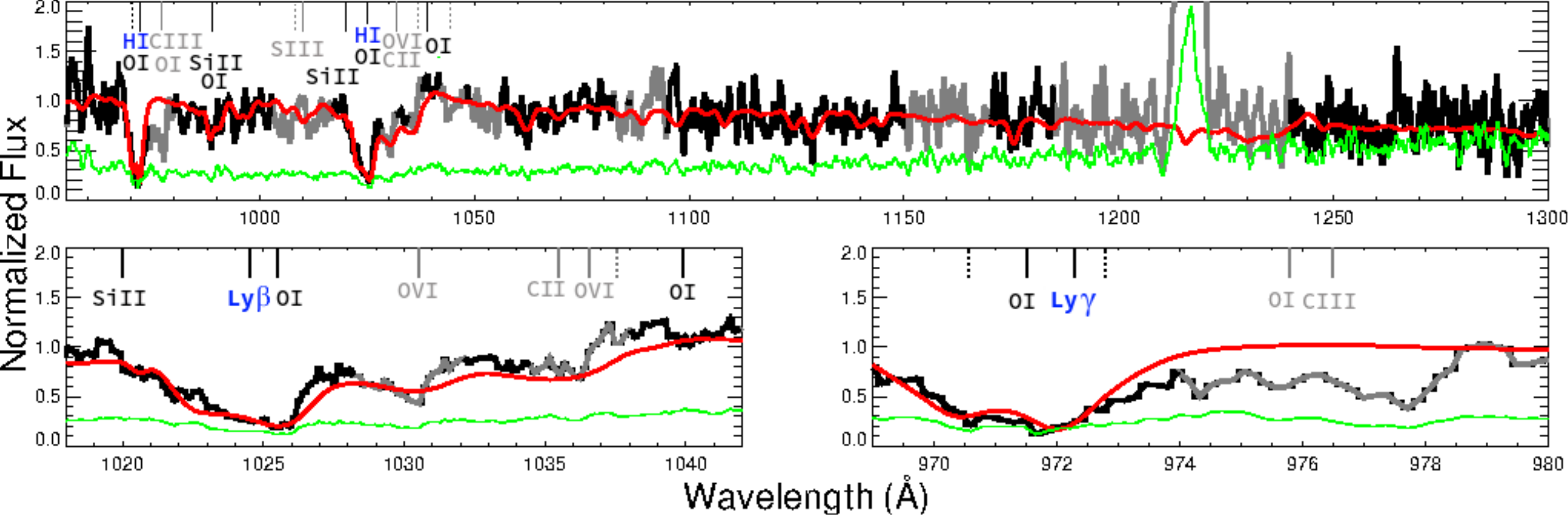}
     \caption{Same as Fig.~\ref{fig:J0921} but for the COS G140L spectrum of J1152+3400. Reference for this observation: \citet{izotov2016a}}
         \label{fig:J1152}
   \end{figure*}

\begin{figure*}
   \centering
   \includegraphics[scale = 0.55]{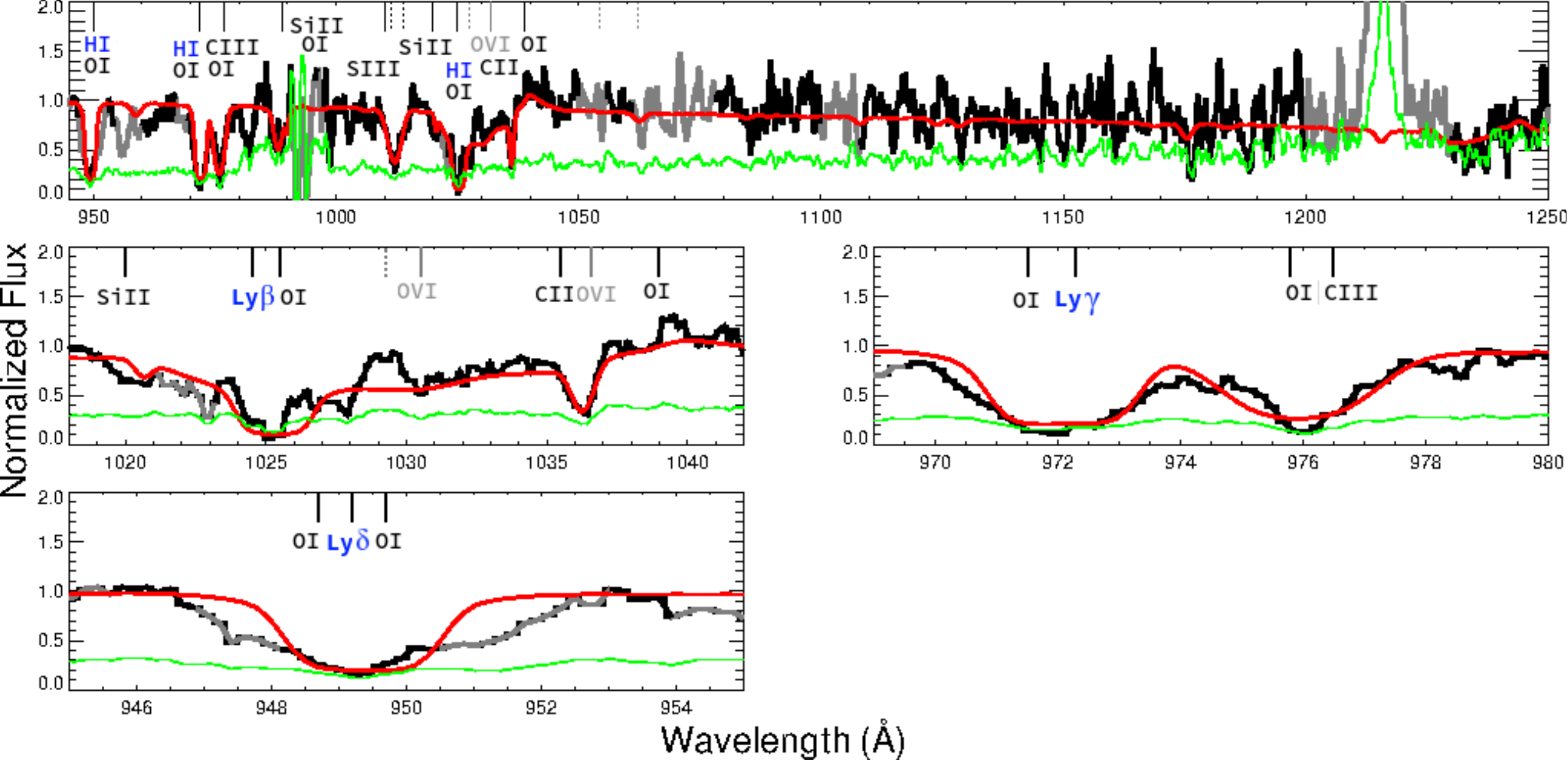}
     \caption{Same as Fig.~\ref{fig:J0921} but for the COS G140L spectrum of  J1333+6246. Reference for this observation: \citet{izotov2016a}}
         \label{fig:J1333}
   \end{figure*}

\begin{figure*}
   \centering
   \includegraphics[scale = 0.55]{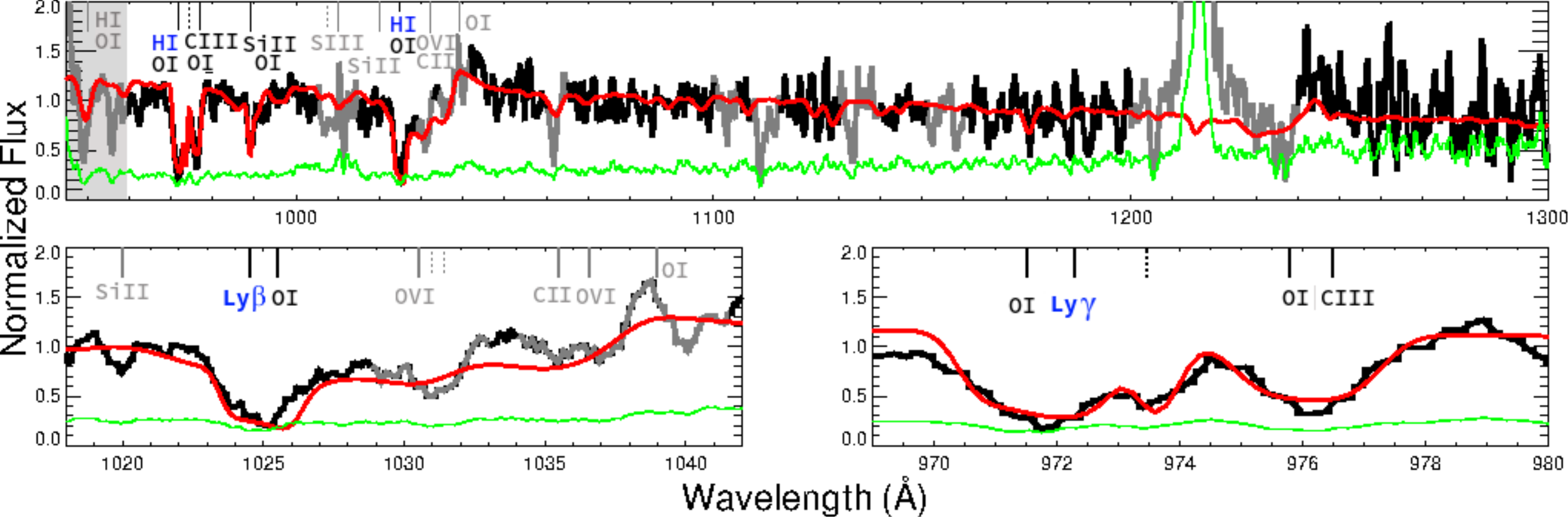}
     \caption{Same as Fig.~\ref{fig:J0921} but for the COS G140L spectrum of J1442-0209. Reference for this observation: \citet{izotov2016a}}
         \label{fig:J1442}
   \end{figure*}

\begin{figure*}
   \centering
   \includegraphics[scale = 0.55]{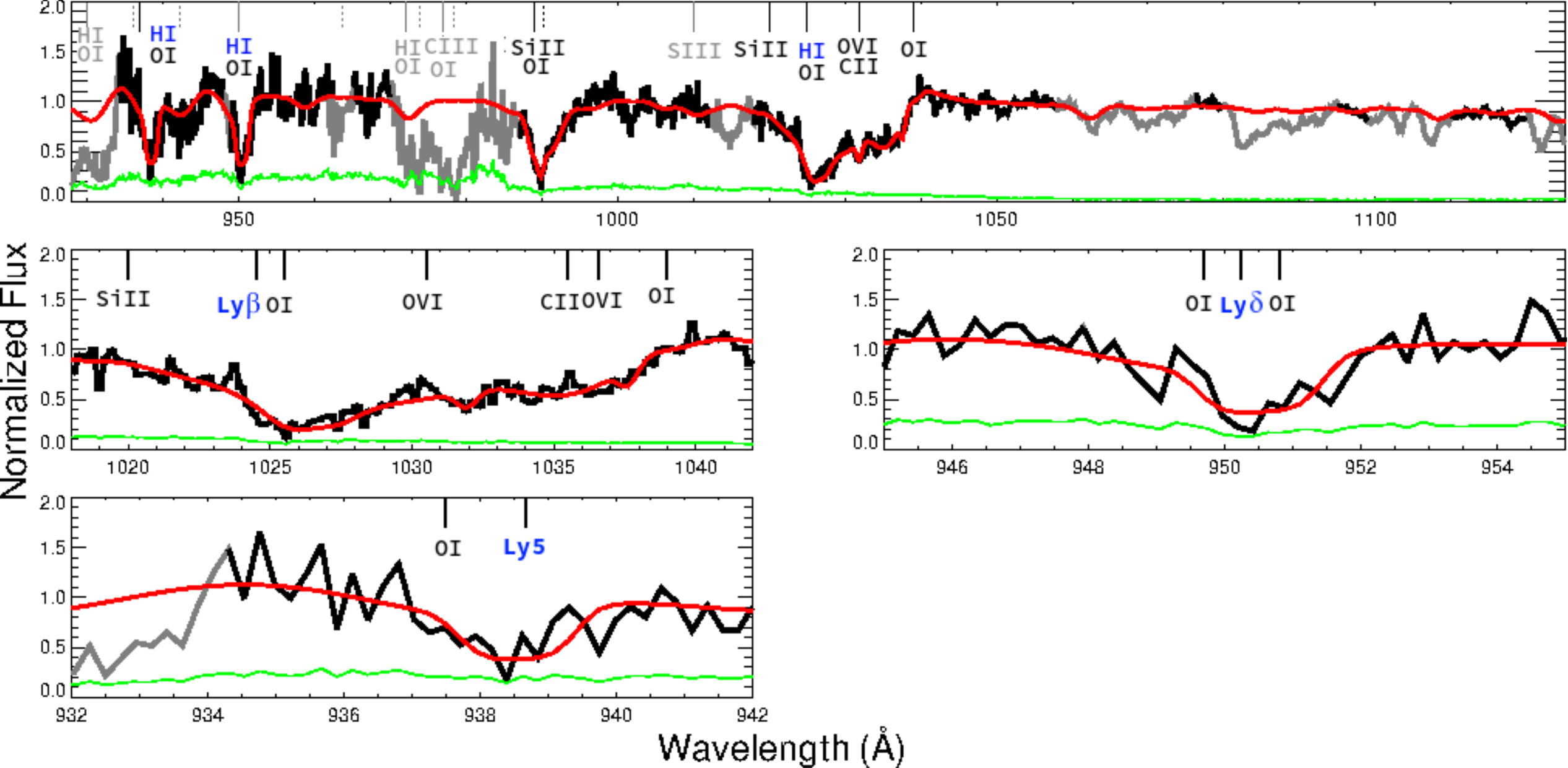}
     \caption{Same as Fig.~\ref{fig:J0921} but for the COS G140L spectrum of  Tol1247-232. Reference for this observation: \citet{leitherer2016}}
         \label{fig:tol1247}
   \end{figure*}

\begin{figure*}
   \centering
   \includegraphics[scale = 0.55]{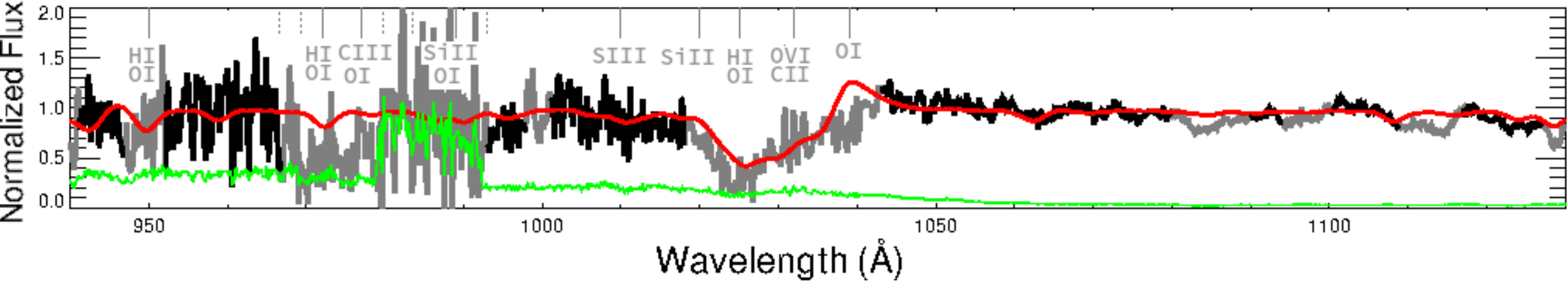}
     \caption{Best fit for the stellar continuum for the COS G140L spectrum of Tol0440-381. This galaxy has a low redshift and the Milky Way absorption lines contaminate the Lyman series. Consequently, we do not fit for the ISM absorption lines. Reference for this observation: \citet{leitherer2016}}
         \label{fig:tol0440}
   \end{figure*}

\begin{figure*}
   \centering
   \includegraphics[scale = 0.55]{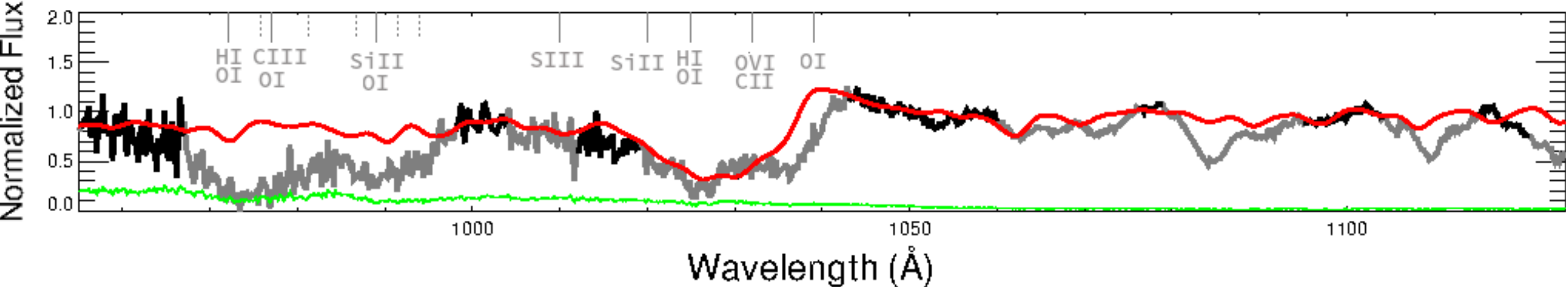}
     \caption{Same as Fig.~\ref{fig:tol0440} but for the COS G140L spectrum of Mrk54. Reference for this observation: \citet{leitherer2016}}
         \label{fig:mrk54}
   \end{figure*}

\begin{figure*}
   \centering
   \includegraphics[scale = 0.55]{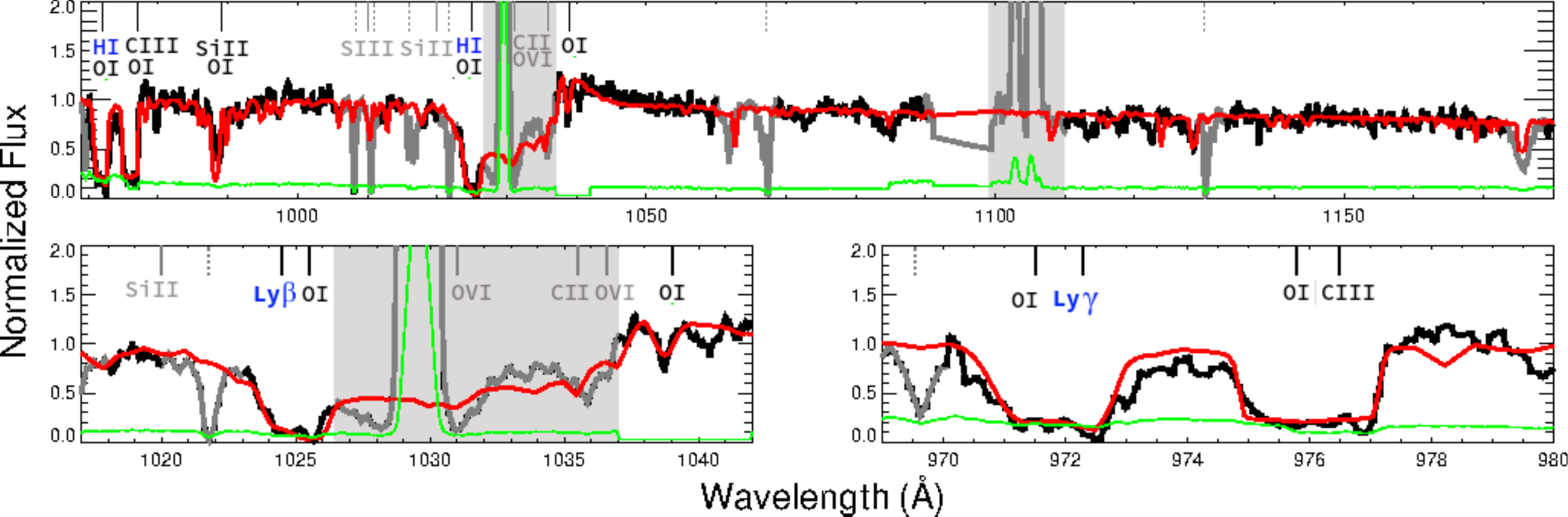}
     \caption{Same as Fig.~\ref{fig:J0921} but for the COS G130M spectrum of J0926+4427. Reference for this observation: \citet{heckman2011}}
         \label{fig:J0926}
   \end{figure*}

\begin{figure*}
   \centering
   \includegraphics[scale = 0.55]{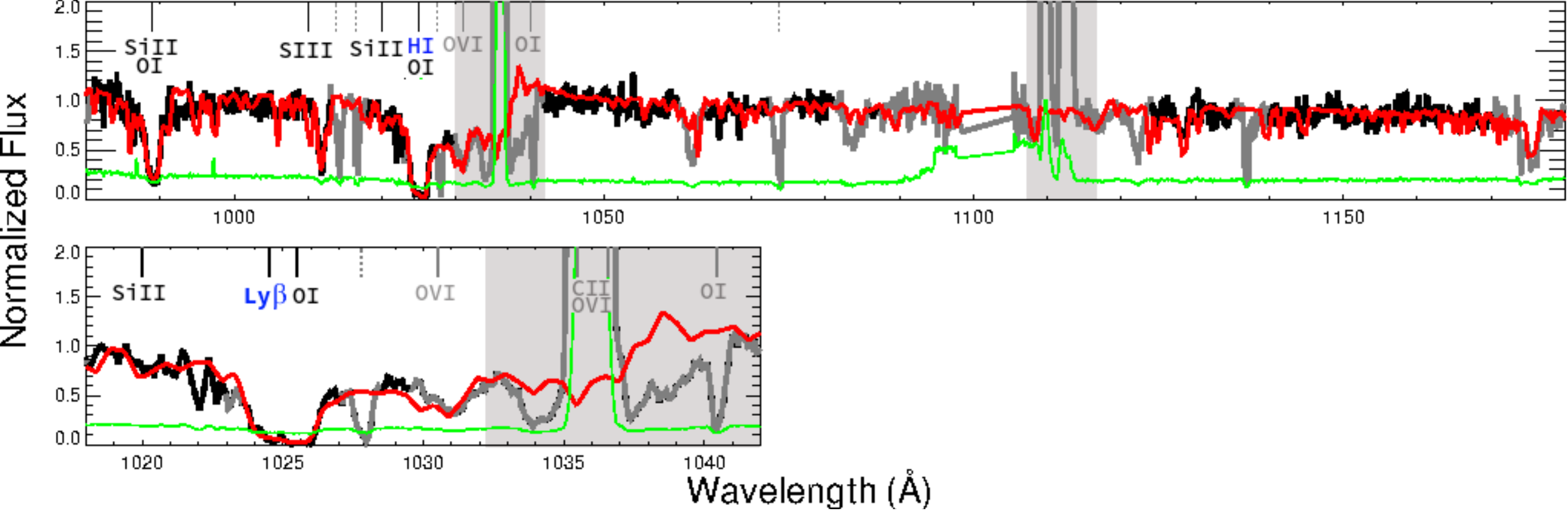}
     \caption{Same as Fig.~\ref{fig:J0921} but for the COS G130M spectrum of J1429+0643. Reference for this observation: \citet{heckman2011}}
         \label{fig:J1429}
   \end{figure*}

\begin{figure*}
   \centering
   \includegraphics[scale = 0.55]{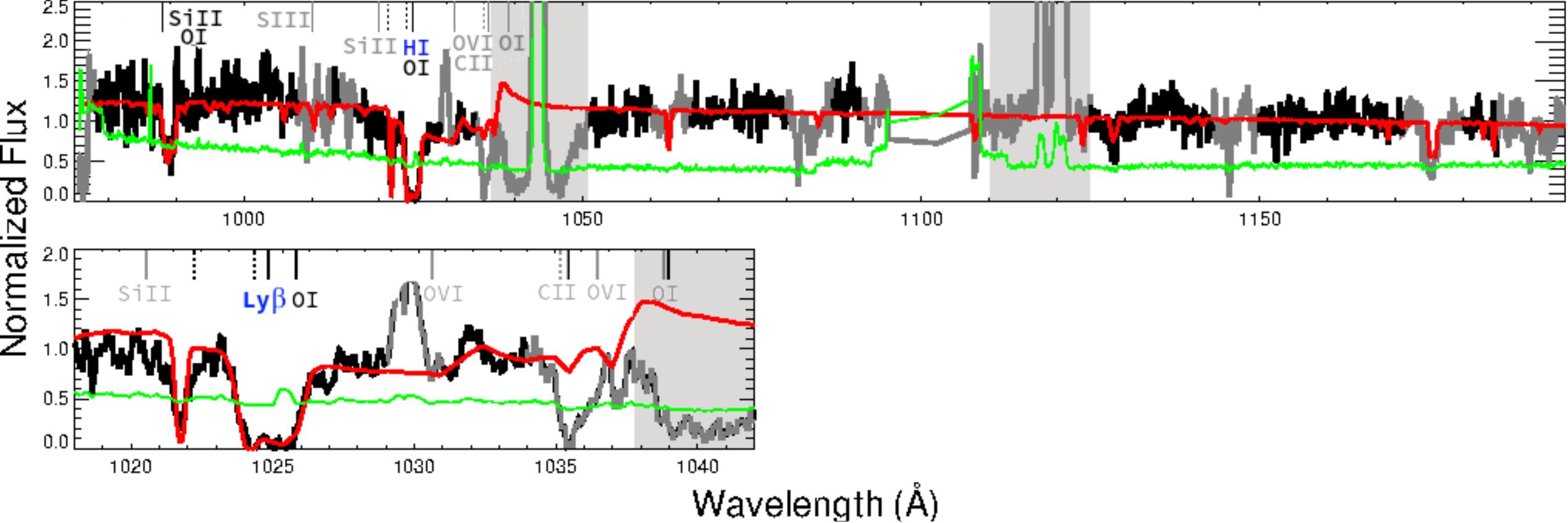}
     \caption{Same as Fig.~\ref{fig:J0921} but for the COS G130M spectrum of GP0303-0759. Reference for this observation: \citet{henry2015}}
         \label{fig:GP0303}
   \end{figure*}

\begin{figure*}
   \centering
   \includegraphics[scale = 0.55]{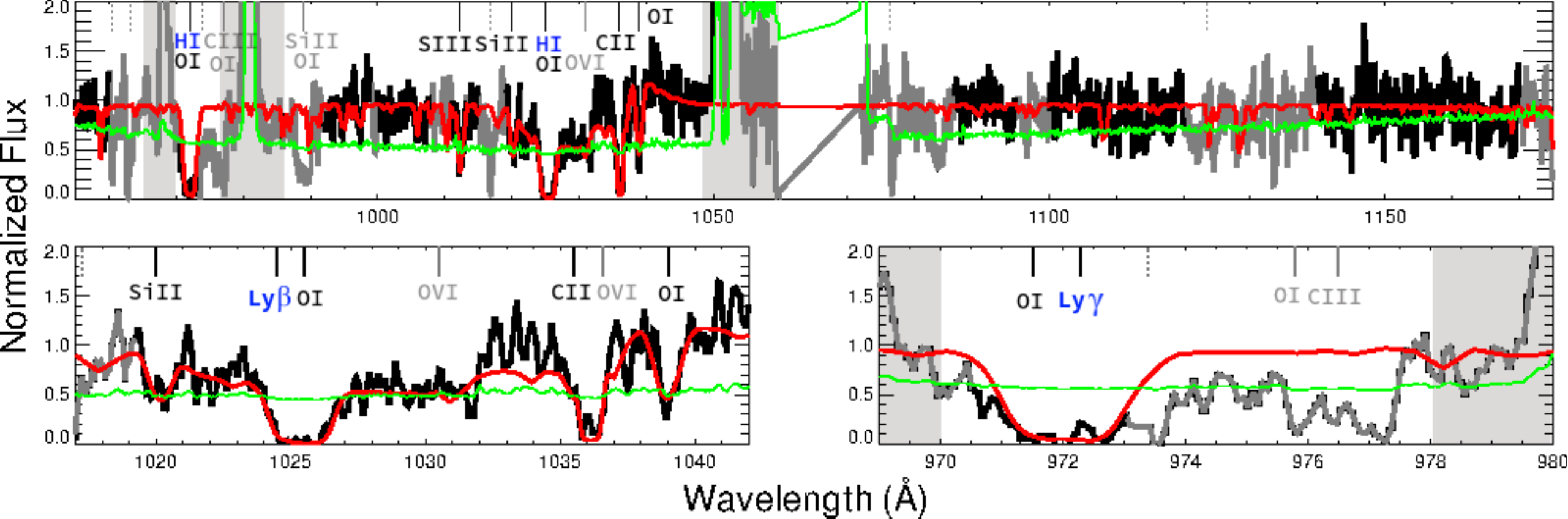}
     \caption{Same as Fig.~\ref{fig:J0921} but for the COS G130M spectrum of GP1244+0216. Reference for this observation: \citet{henry2015}}
         \label{fig:GP1244}
   \end{figure*}

\begin{figure*}
   \centering
   \includegraphics[scale = 0.55]{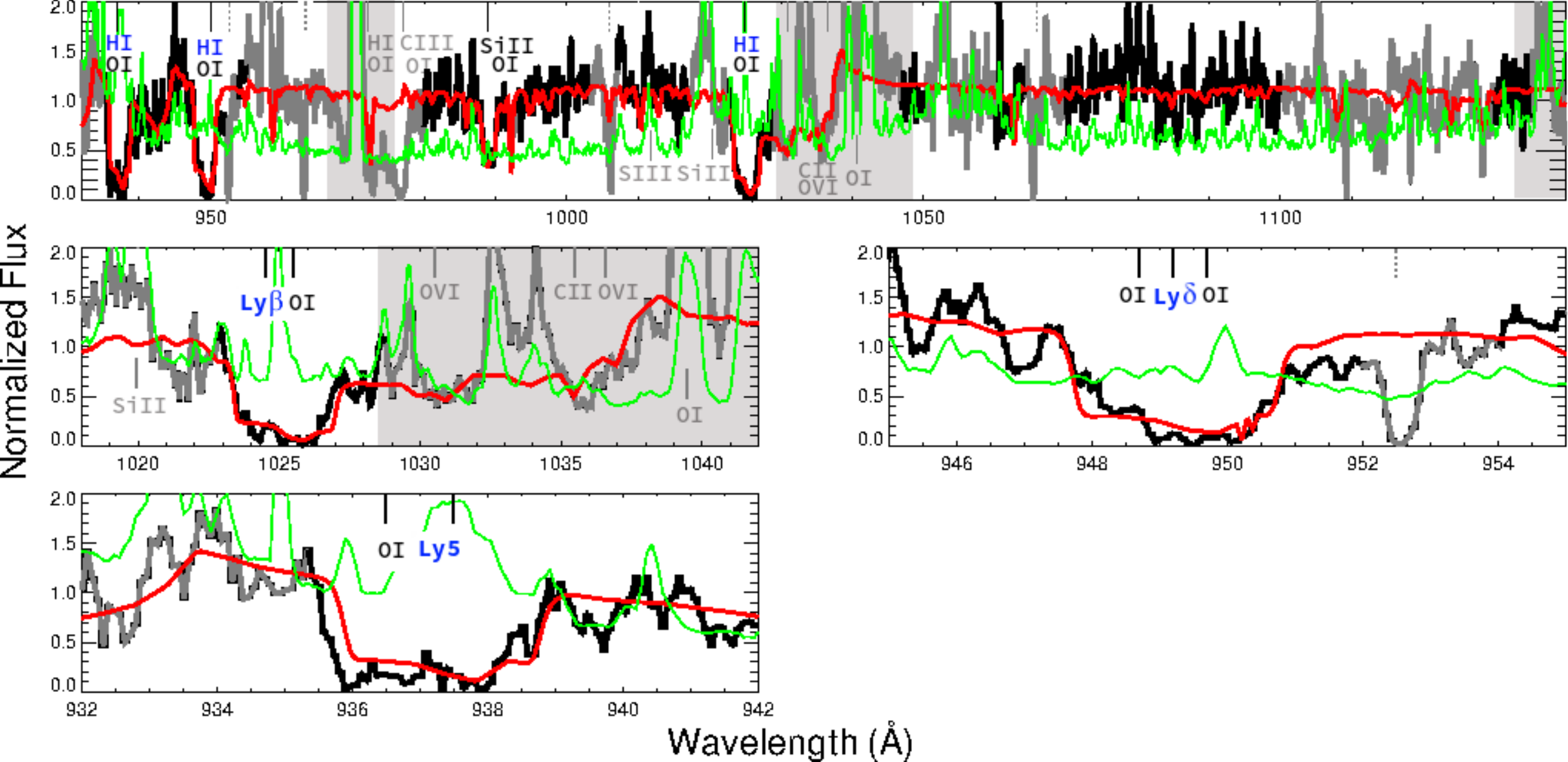}
     \caption{Same as Fig.~\ref{fig:J0921} but for the COS G130M spectrum of  GP1054+5238. Reference for this observation: \citet{henry2015}}
         \label{fig:GP1054}
   \end{figure*}

\begin{figure*}
   \centering
   \includegraphics[scale = 0.55]{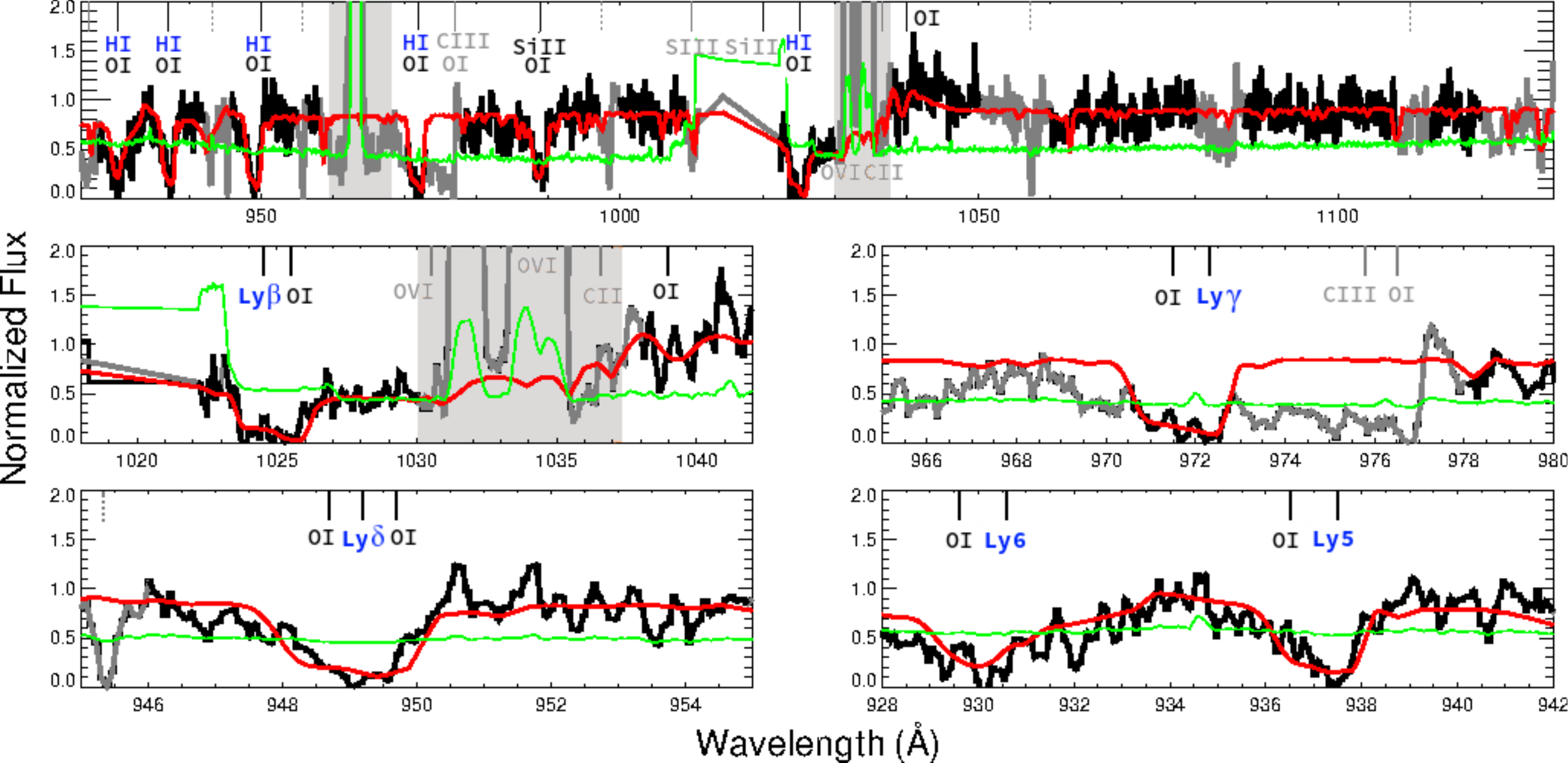}
     \caption{Same as Fig.~\ref{fig:J0921}  but for the COS G130M spectrum of GP0911+1831. Reference for this observation: \citet{henry2015}}
         \label{fig:GP0911}
   \end{figure*}

\begin{figure*}
   \centering
   \includegraphics[scale = 0.55]{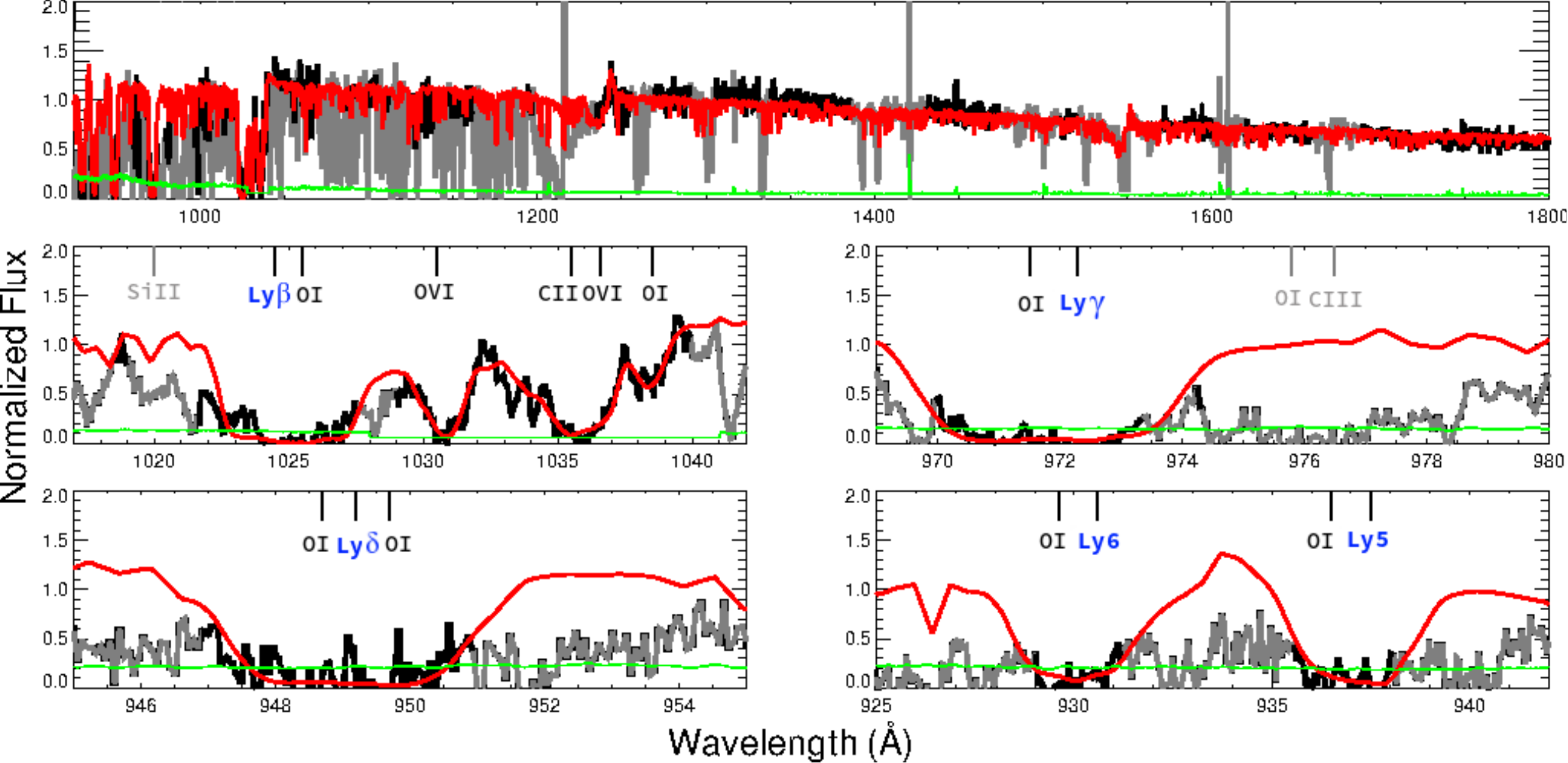}
     \caption{Same as Fig.~\ref{fig:J0921} but for the MagE spectrum of  SGAS J122651.3+215220. Reference for this observation: \citet{rigby}}
         \label{fig:S1226}
   \end{figure*}

\begin{figure*}
   \centering
   \includegraphics[scale = 0.55]{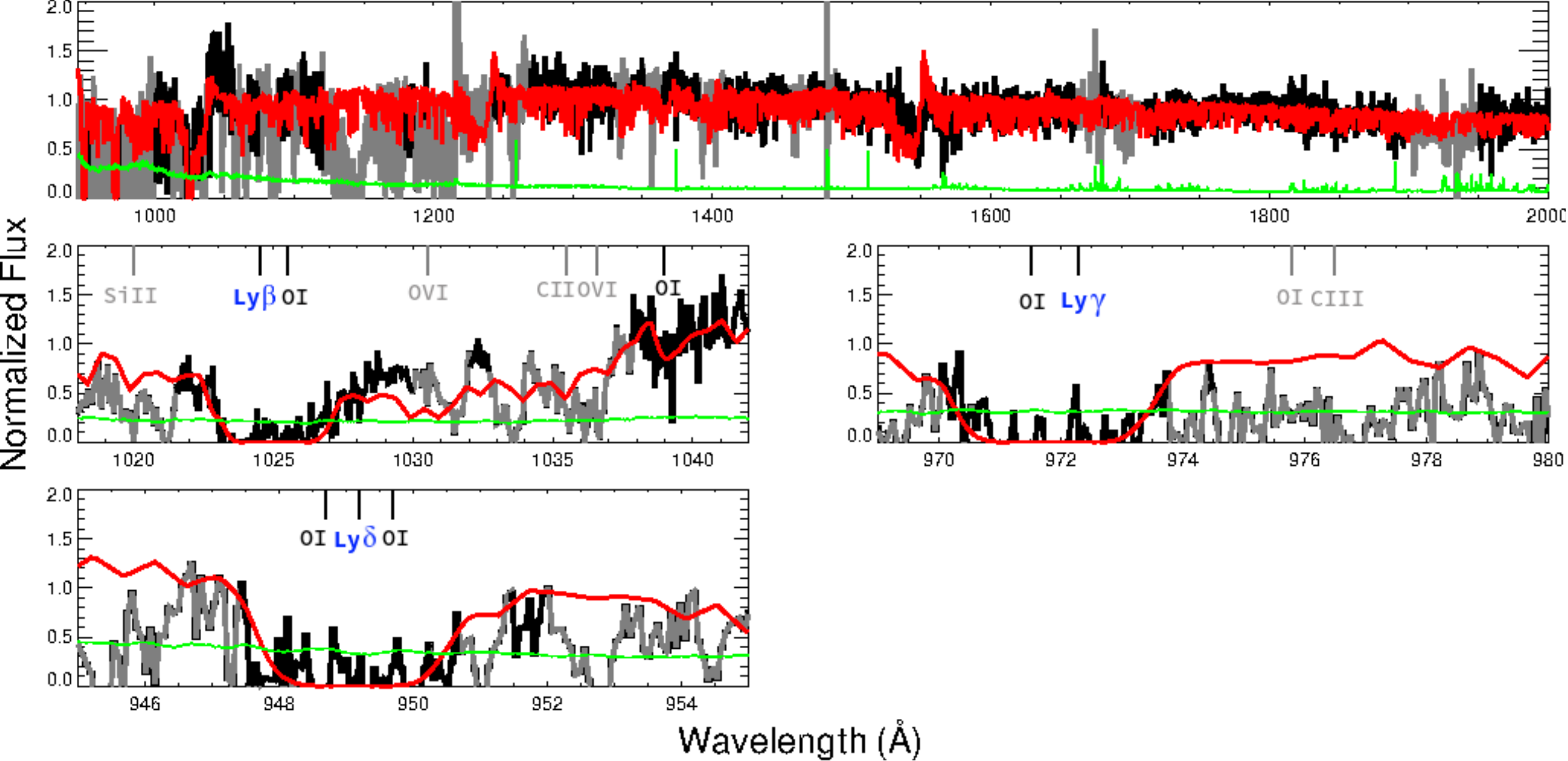}
     \caption{Same as Fig.~\ref{fig:J0921} but for the MagE spectrum of SGAS J152745.1+065219. Reference for this observation: \citet{rigby}}
         \label{fig:S1527}
   \end{figure*}

\begin{figure*}
   \centering
   \includegraphics[scale = 0.55]{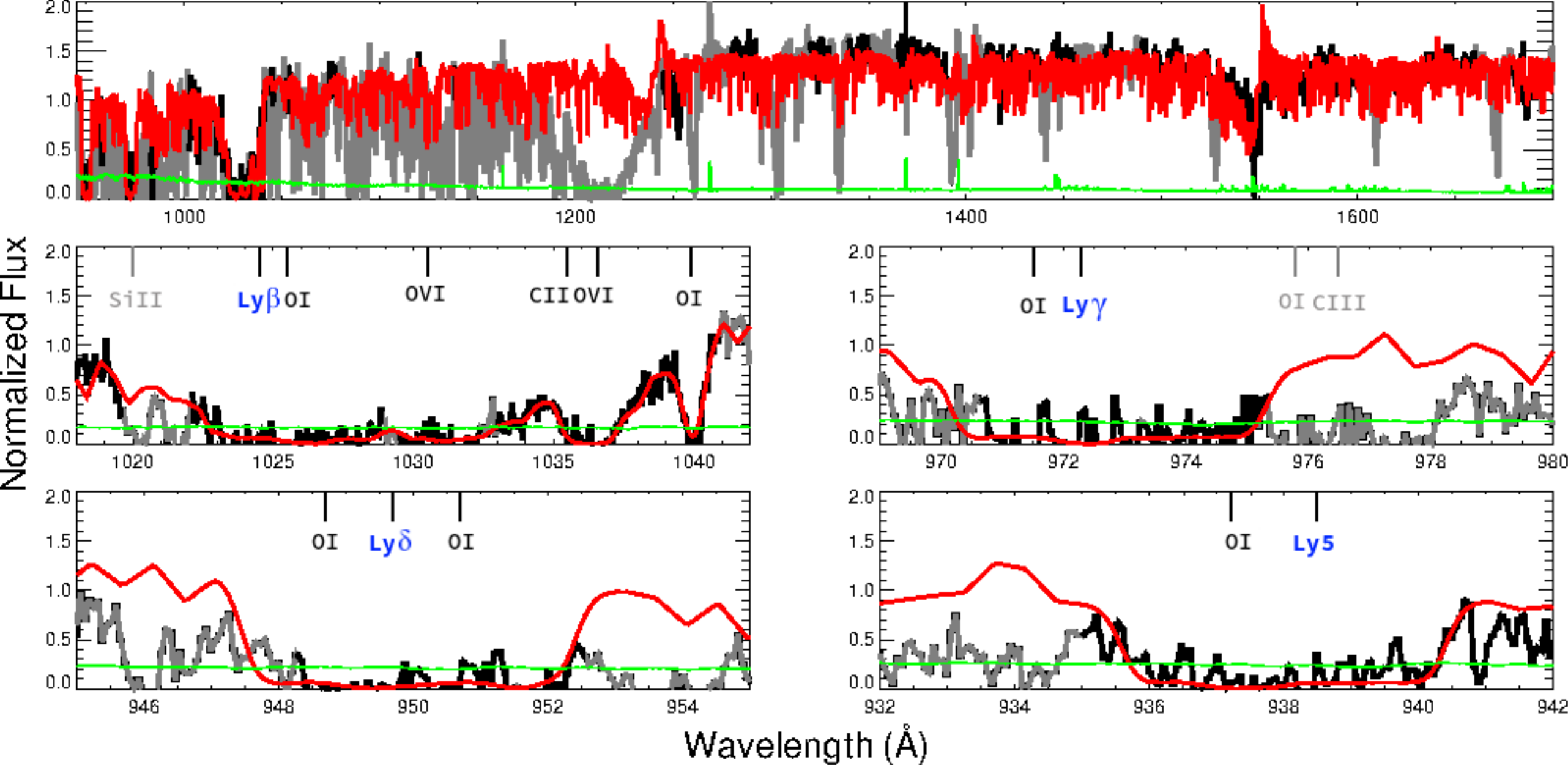}
     \caption{Same as Fig.~\ref{fig:J0921} but for the MagE spectrum of the Cosmic Eye. Reference for this observation: \citet{rigby}}
         \label{fig:cosmic}
   \end{figure*}

We provide fits of the synthetic and observed spectra detailed in this article. Fig.~\ref{fig:singlefit15000} and Fig.~\ref{fig:singlefit600} show the synthetic spectra for three S/N ratios ($\infty$, 10 and 2) with R~=~15000 and  R~=~600, respectively. In Figs~\ref{fig:J0921}--\ref{fig:cosmic} we present the fits for the 18 galaxies in our sample. The top panels of each figure show the full wavelength coverage; the observed flux is indicated in black and the total fit in red. Gray flux indicates that these regions were masked during the fit. The green lines show the error on the observed flux. The lower panels zoom in on individual Lyman series lines. The upper portions of each plot indicate the fitted Lyman series lines (blue), ISM metal absorption lines (solid black line), and Milky Way absorption lines (black dashed lines). The line labels are gray if they are not fit. Geocoronal regions are denoted by shaded gray regions. 

\label{app:fits}

\clearpage  
\section{Tables}

\subsection{Measured systematic errors from single synthetic spectra}

We tabulate the systematic errors measured from the synthetic spectra (Sect.~\ref{sect:sims}). Table~\ref{table:density} gives the $N_\ion{H}{i}$ systematic errors, and Table~\ref{table:picket} gives the $C_f$ systematic errors. The percent errors plotted in Fig.~\ref{figure:picket} are computed by subtracting 0.9 from the errors and dividing the difference by 0.9 (the $C_f$ used to generate the spectra).

    \begin{table}[!htbp]
    \centering          
    \caption{Systematic errors on the logarithm of the \ion{H}{i} column density estimated from synthetic spectra.}            
    \label{table:density}  
    \begin{tabular}{c c c c c c c c }    
    \hline\hline       
        (1) &  (2) & (3) & (4) & (5) &  (6) & (7) & (8)  \\  
    \multirow{1}{*}{R}& \multicolumn{7}{c}{Signal-to-noise ratio} \\ 
     &  50 & 40 & 30 &20 &10& 5&2 \\ \hline 
    15000 & 0.04 & 0.04 & 0.05 & 0.07 & 0.17 & 0.33 & 1.13 \\ 
    6000 &  0.05 & 0.07 & 0.10 & 0.14&  0.31 & 0.78 &  1.27  \\ 
    3000 & 0.09 &  0.09 &  0.12& 0.18 & 0.57 & 1.05 &  1.13  \\ 
    1500 & 0.10 & 0.17 & 0.20 &  0.28 &  0.65 &  1.04 &  1.28  \\
    1000   & 0.12 &  0.14 &  0.17 &  0.30 &  0.87 &  1.21 &  1.18  \\
    750   &  0.14 &  0.17 &  0.24 & 0.33 &  0.86 &  1.20 &  1.53 \\
    600  &  0.13 &  0.19 &  0.25 & 0.30 & 0.81 &  1.21 &  1.54  \\
    \hline   
    \end{tabular}
        \tablefoot{Columns 2-8 give the systematic log($N_\ion{H}{i}$) error at various spectral resolutions (R; see column 1) and S/N ratios (see columns 2-7 for S/N 50-2) for simulated spectra with $N_\ion{H}{i} = 10^{17.57}$cm$^{-2}$. See Sect.~\ref{simcold}.}
    \end{table}

    \begin{table}[!htbp]
    \centering          
        \caption{Systematic errors on the covering fraction estimated from synthetic spectra.}
    \label{table:picket}  
    \begin{tabular}{c c c c c c c c }     
    \hline\hline       
        (1) &  (2) & (3) & (4) & (5) &  (6) & (7) & (8)  \\  
    \multirow{1}{*}{R}& \multicolumn{7}{c}{Signal-to-noise ratio} \\ 
     &  50 & 40 & 30 &20 &10& 5&2 \\ \hline

    15000 & <0.01 & <0.01 & <0.01 &  0.01 & 0.01&  0.02&  0.06 \\ 
    6000  & <0.01 & <0.01 & <0.01 &  0.01&  0.01&   0.03 &  0.07 \\ 
    3000  & <0.01 & <0.01 & <0.01&  0.01 &  0.01 & 0.03 & 0.07  \\ 
    1500  & <0.01 &  0.01 &  0.01 &  0.01  &  0.03 &  0.05 &  0.10 \\
    1000  &  0.01 & 0.011 &  0.02 & 0.02    &  0.04 &  0.07 & 0.12 \\
    750   &  0.02 &  0.021 &  0.03 & 0.04   &  0.06 & 0.09 &  0.14 \\
    600   &  0.03 & 0.032 &  0.04 &  0.05   &  0.09 &  0.10 & 0.17 \\
    \hline                  
    \end{tabular}
        \tablefoot{Columns 2-8 give the systematic $C_f$ errors at different spectral resolutions (R; column 1) and S/N (see columns 2-7 for S/N 50-2) for simulated spectra with $C_f = 0.9$.  See Sect.~\ref{simcf}.}
    \end{table}

\subsection{Lyman series residual flux}

Table~\ref{table:cfdepth} lists the residual flux of the individual Lyman series lines (Sect.~\ref{method:depth}).

 \begin{table*}[!htbp]
    \caption{Measurement of \ion{H}{i} covering fraction derived from the residual flux of the individual Lyman series. See Sect.~\ref{method:depth} }
     \label{table:cfdepth}
    \centering   
    \begin{tabular}{c c c c c c}
    \hline \hline
    Galaxy name&   Ly$\beta$ & Ly$\gamma$ & Ly$\delta$ & Ly5  & Weighted mean \\ \hline 
    J0921+4509  &  0.769 $\pm$ 0.116 & - & - & - & 0.769 $\pm$ 0.116 \\ 
    J1503+3644  &  0.847 $\pm$ 0.157 & 0.723 $\pm$ 0.129 & 0.741 $\pm$ 0.112 & 0.744 $\pm$ 0.113 & 0.754 $\pm$ 0.062  \\
    J0925+1409 &  0.635 $\pm$ 0.094 & - & - & - & 0.635 $\pm$ 0.094  \\ 
    J1152+3400  & 0.619 $\pm$ 0.088 & - & - & - & 0.619 $\pm$ 0.088  \\ 
    J1333+6246  & 0.773 $\pm$ 0.152 & 0.870 $\pm$ 0.103 & 0.804 $\pm$ 0.106 & - & 0.826 $\pm$ 0.066  \\
    J1442-0209  &  0.589 $\pm$ 0.049 & 0.471 $\pm$ 0.068 & - &  - & 0.549 $\pm$ 0.040 \\ 
    Tol1247-232 & 0.543 $\pm$ 0.135 & - & 0.775 $\pm$ 0.134  &  0.791 $\pm$ 0.175  & 0.690 $\pm$ 0.084 \\ Tol0440-381 & 0.507 $\pm$ 0.139 & - & 0.615 $\pm$ 0.167  &  0.602 $\pm$ 0.137  & 0.570 $\pm$ 0.084 \\
    Mrk54      & 0.397 $\pm$ 0.136 & - & 0.652 $\pm$ 0.120  &  0.406 $\pm$ 0.148  & 0.504 $\pm$ 0.077 \\
    J0926+4427  &  0.817 $\pm$ 0.057 & 0.807 $\pm$ 0.087 &  - & -  & 0.814 $\pm$ 0.048 \\ 
    J1429+0643  &  0.955 $\pm$ 0.061 & - & - & - & 0.955 $\pm$ 0.061 \\ 
    GP0303-0759  & - & - & - & - & -\\ 
    GP1244+0216  &  0.985 $\pm$ 0.211 & 0.894 $\pm$ 0.204 & 1.000 $\pm$ 0.292 & -  & 0.950  $\pm$ 0.131 \\ 
    GP1054+5238  & 0.936 $\pm$ 0.318 & - & 0.891 $\pm$ 0.203 & 0.798 $\pm$ 0.420  & 0.889  $\pm$ 0.158 \\ 
    GP0911+1831  &  0.718 $\pm$ 0.361 & 0.781 $\pm$ 0.189 & 0.825 $\pm$ 0.198 & 0.635 $\pm$ 0.282 & 0.765  $\pm$ 0.116\\
    \sone & 1.000 $\pm$ 0.010 & 1.000 $\pm$ 0.030 & 0.981 $\pm$ 0.057 & 0.940 $\pm$ 0.060  & 0.998 $\pm$ 0.009  \\ 
     \stwo &   0.858 $\pm$ 0.141  & 1.000 $\pm$ 0.045 & 1.000 $\pm$ 0.087 & 1.000 $\pm$ 0.201 & 0.990 $\pm$ 0.038 \\
    Cosmic Eye &  1.000$\pm$ 0.043 & 1.000 $\pm$ 0.063 & 1.000 $\pm$ 0.033 & 0.899 $\pm$ 0.166 & 0.998 $\pm$ 0.024 \\ \hline     
    \end{tabular}
    \tablefoot{Dashes indicate that these transitions were not observed owing to Milky Way absorption, geocoronal emission, or low S/N.}
    \end{table*}
 
\end{appendix}

\end{document}